\documentclass[]{biophys-new}
\usepackage[utf8]{inputenc}
\usepackage{graphicx}
\usepackage[colorlinks,allcolors=cyan!70!black]{hyperref}

\usepackage{bm}
\usepackage{amsmath}
\usepackage{amssymb}
\usepackage{subfigure}
\usepackage{url}
\runningtitle{Folding analysis by persistent homology}
\runningauthor{Ichinomiya, Obayashi and Hiraoka}

\title{Protein folding analysis using features obtained by persistent homology}

\author[1,2,*]{Takashi Ichinomiya}
    \corrauthor[*]{tk1miya@gifu-u.ac.jp}
    \affil[1]{Gifu University School of Medicine, Yangido 1-1, Gifu, 501-1194, Gifu, Japan}
    \affil[2]{The United Graduate School of Drug Discovery and Medical Information Sciences of Gifu University, Yanagido 1-1, Gifu, 501-1194, Gifu, Japan }
\author[3]{Ippei Obayashi}
\affil[3]{Center for Advanced Intelligence Project, RIKEN, Nihonbashi 1-chome Mitsui Building, 15th floor,1-4-1 Nihonbashi, Chuo-ku, Tokyo 103-0027, Japan}
\author[4,3]{Yasuaki Hiraoka}%
\affil[4]{%
WPI-ASHBi, Kyoto University Institute for Advanced Study, Kyoto University, Yoshida Ushinomiya-cho, Sakyo-ku, Kyoto 606-8501, Japan }

\date{\today}

\begin{document}
 \begin{frontmatter}
  \begin{abstract}
Understanding the protein folding process is an outstanding issue in biophysics; recent developments in molecular dynamics simulation have provided insights into this phenomenon. However, the large freedom of atomic motion hinders the understanding of this process. In this study, we applied persistent homology, an emerging methods to analyze topological features in a dataset, to reveal protein folding dynamics. We developed a new method to characterize protein structure based on persistent homology  and applied this method to molecular dynamics simulations of chignolin. Using principle component analysis or non-negative matrix factorization, our analysis method revealed two stable states and one saddle state, corresponding to the native, misfolded, and transition states, respectively. We also identified an unfolded state with slow dynamics in the reduced space. Our method serves as a promising tool to understand the protein folding process.
  \end{abstract}
  
  \begin{sigstatement}
  To understand the protein folding process, protein forms  must be presented in a comprehensible way. In this paper, we propose a method to represent the internal protein configuration using persistent homology, an emerging data analysis technique based on topology. Using this method, we simplified the complex dynamics of chignolin and  identified two metastable and transition states as fixed points. Our method is applicable to other macromolecules and will help to understand the functions and dynamics of biomolecules such as proteins and DNA.
  \end{sigstatement}
 \end{frontmatter}
\maketitle


\section{Introduction}\label{Sec:Introduction}

Since the proposal of Levinthal's paradox in 1968, the folding of biomolecules, including proteins, has attracted the interest of numerous scientists\cite{Levinthal1968AreFolding}. Molecular dynamic (MD) simulations have contributed to the understanding of the folding  mechanisms\cite{Lane2013}. 
However, the atoms in the MD simulations have a large degree of freedom, and the essential folding dynamics must be extracted to comprehend the protein dynamics. Therefore, several methods have been proposed, such as principal component analysis(PCA)\cite{Kitao1991TheVacuum}, relaxation mode analysis(RMA)\cite{doi:10.1063/1.4931813},  time structure-based independent component analysis(tICA)\cite{doi:10.1063/1.3554380, Schwantes2016}, and manifold-learning\cite{Fujisaki2018}.

Previous studies attempted to identify the essential motion related to the large deformation that leads to protein folding. However, the definition of "large deformation" is ambiguous. For example, when a protein unfolds into nearly a straight line, a small bend at the center of the molecule will cause a large dislocation of the atoms at the end of the chain. In this case, the deformation in a Ramachandran plot\cite{Ramachandran1968} is small, but large in atom Cartegian coordinates. Moreover, the importance of deformations also depends on the protein  structure. For example, in a small protein that has only one $\beta$-sheet, a small change in the bond angle at the hairpin of the molecule may disrupt the $\beta$-sheet structure.  Thus, this small change in the angle results in a "large deformation". Alternatively, if this protein is completely unfolded, then a slight change in the hairpin region bond angle does not cause a "large deformation". These examples show the difficulties in defining a  "large deformation" in a protein.

We propose using topological data analysis(TDA) to characterize the structure and deformation of a protein. Using TDA, we investigated the topological signatures  such as loops or vacancies, embedded in a dataset. This approach yields  successful results in  many fields, including  RNA hairpin folding analysis\cite{doi:10.1063/1.3103496} or gene regulation networks\cite{Nicolau2011}. TDA has several advantages compared with  standard protein structure analysis tools, such as Ramachandran plots, distance matrices, and the atomic cartesian coordinates. First, TDA captures  changes in the global structure, whereas other methods, such as Ramachandran plots only consider local properties, such as bond angles. In contrast, "loops" or "vacancies" are formed by several atoms. Thus, TDA captures the non-local structure. Second, topological changes strongly depend on the atom conformation. For example, if a protein forms a straight chain, then there are no loops. If a small bend occurs at the center of this chain, then the atoms at the end of the chain exhibit large dislocations; however, loops do not form. Alternatively, a small change in the bond angle at the hairpin of a $\beta$-sheet can break the loops formed by the atoms in the $\beta$-sheet. Finally, TDA provides intuitive insights into  protein dynamics. For example, loop emergence and disappearance are  more clearly visualized using TDA than using the coordinated atomic motion. 

Here we applied persistent homology (PH) analysis \cite{Edelsbrunner2002} for TDA. PH is based on algebraic topology,  and has been applied to many problems in physics, chemistry, biology, and medicine\cite{Xia2014, Xia2015MultidimensionalData, Carlsson2014, Hiraoka2016,Ichinomiya2017}. Although PH is a highly effective tool for the analysis of non-local structures, it has several inherent limitations. First, PH results are sometimes difficult to interpret. In the original PH analysis, we obtain two values called "birth" and "death" for each loop or cavity, and make decisions based on the distribution of these values. Frequently, these two values are insufficient to understand the physical relevance provided by PH. For example, consider the folding of a protein which has two $\alpha$-helices. If the birth and death values obtained from these $\alpha$-helices are nearly identical, it is difficult to distinguish which $\alpha$-helix is created first in the folding process. Recently, Escolar {\it et al.} developed a method to calculate "volume optimal cycles," which enables identification of the atoms that form loops or cavities\cite{Escolar2016, Obayashi2018}. This method is useful to explain PH results and has revealed hidden structures in glass and amorphous polymers\cite{Hiraoka2016, Ichinomiya2017}. Another difficulty of PH lies in the fluctuation in the loop number. Even if the number of atoms is constant, the number of loops obtained by PH depends on the configuration of the atoms. However, standard machine-learning techniques, such as PCA or k-means clustering, require that all the input data have the same dimension.  These machine-learning techniques were avoided in previous studies using PH to analyze biomolecular structures\cite{Xia2014,Xia2015MultidimensionalData}. 
To overcome this difficulty, several methods such as persistent diagram vectorization\cite{Kimura2018}, kernel methods\cite{Kusano2016}, and persistent landscapes\cite{Bubenik2014} have been proposed.

In this paper, we propose a new technique to apply machine learning to PH analysis. The key concept is to construct a "topological feature vector"(TFV) using volume optimal cycles. In this approach, we considered the volume optimal cycles as the "text" that describes the protein structure. Each volume optimal cycle is a collection of simplices (edges or faces), similar to a text being a collection of words. This concept enables the use of text-mining techniques. Next, we applied PCA and non-negative matrix factorization (NMF) to reduce the TFVs obtained from MD simulations of chignolin. Finally, we compared the result with analyses based on atom-position and contact mapping. A previous study showed that chignolin has native, misfolded, unfolded, and intermediate structures \cite{doi:10.1063/1.4931813, Fujisaki2018}. Therefore, we performed a full atomic MD simulation of chignolin in aqueous solution, and analyzed the result using TFV. We observed that NMF of TFV provides essential information on protein structure and dynamics. Additionally, we found that the dynamics in the reduced space yielded two stable- and one saddle-fixed points, which correspond to native, misfolded, and transition states, respectively. The unfolded state did not corresponds to a fixed point. However, the dynamics in the unfolded state were extremely slow.

The remainder of the paper is structured as follows: In Sec.~\ref{Sec:Method}, we describe the PH method, TFV construction, and dimension reduction by NMF. We also describe the details of the chignolin MD simulation. In Sec.~\ref{Sec:Result}, we present the analysis results and compare them with analysis based on cartesian coordinates and contact mapping. PH provides an  intuitive description of the folded, misfolded, transition, and unfolded states. The challenges to overcome, as well as the  future direction are discussed in Sec.~\ref{Sec:Conclusion}.

\section{Method}\label{Sec:Method}

Our analysis process is composed of three procedures. First, we performed PH analysis and identified all loops with their volume optimal cycles. Second, we constructed a "TFV", which  stores the edge contributions to the volume optimal cycles formation are stored. Third, we reduce the dataset dimensionality using PCA or NMF. We explain each step in the following sections. The dataset and scripts we used are uploaded on Open Science Framework (https://osf.io/hsp5w).

\subsection{Persistent homology with volume optimal cycles}

The general mathematical definition of PH is described in terms of the filtration of simplicial complexes\cite{Edelsbrunner2002} or quiver representation\cite{Zomorodian2005}. 
In this section, we explain the degree 1 PH of an $\alpha-$complex composed of a point cloud, which was used to analyze protein folding. 

\begin{figure}
    \centering
    \includegraphics[width = \textwidth]{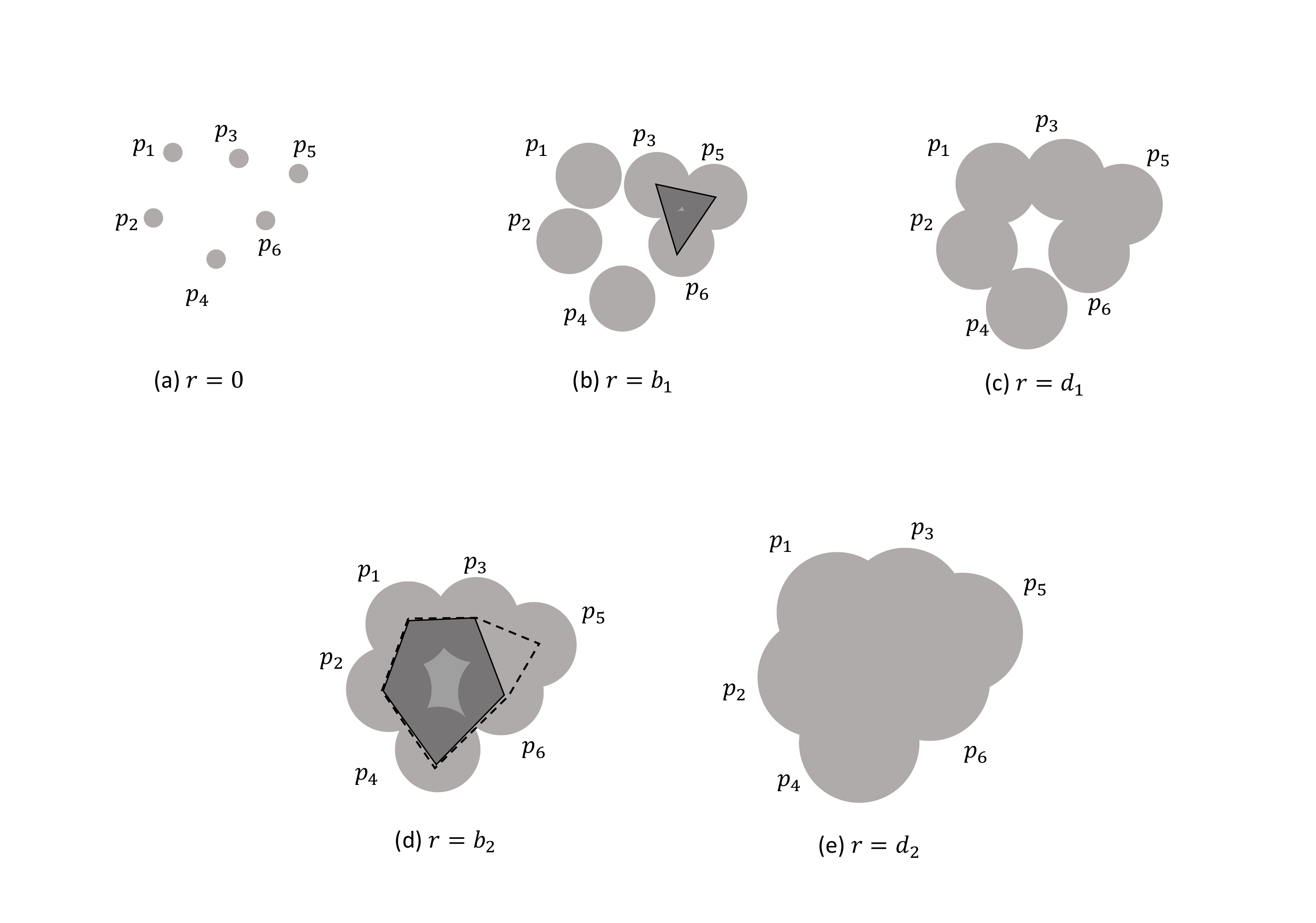}
    \caption{Example of persistent homology analysis. If the radii of the balls is 0, then all atoms are disconnected (a). As we increase the size, the balls coalesce. At $r = b_1$, we obtain a loop surrounded by three edges, $(p_3 p_5)$, $(p_3 p_6)$, and $(p_5 p_6)$ (b). This loop is destroyed when we increase the size of the balls to $r=d_1$ (c). If we increase $r$ further, a new loop appears (d). In this case, we can take several sets of edges that surround the empty space, depicted as solid and dashed lines. In this case, the volume optimal cycle is $(p_1 p_2), (p_2 p_4), (p_4 p_6), (p_6 p_3)$ and $(p_3 p_1)$, depicted by solid lines. This loop is destroyed at $r=d_2$ (e).}
    \label{fig:schematic}
\end{figure}

Consider there are $n$ atoms at $p_1 = (x_1, y_1, z_1), p_2 = (x_2, y_2, z_2), \cdots, p_n= (x_n, y_n, z_n)$  in a three-dimensional space (Fig.~\ref{fig:schematic}). The PH of the $\alpha$-complexes can be regarded as a topological structure when we place a ball of radius $r$ at $p_1, p_2,\cdots, p_n$. If $r=0$, all the balls are disconnected  (Fig.~\ref{fig:schematic}(a)). As we increase $r$, the balls coalesce, and a loop emerges at $r=b_1$ (Fig. \ref{fig:schematic}(b)). We call $b_1$  the "birth" of this loop, and  the three edges, $(p_3 p_5)$, $(p_3 p_6)$, and $(p_5 p_6)$, surround this loop. This loop shrinks as $r$ increases, and at $r=d_1$, the loop is fulfilled and disappears (Fig.~\ref{fig:schematic}(c)). We call  $d_1$ the "death" of this loop. In this case, the edges that surround the loop are unique, however, not always uniquely determined. For example, in the case of Fig.~\ref{fig:schematic}(d),  the loop that emerged at $r= b_2$ is surrounded by 5 edges, $(p_1 p_2), (p_2 p_4), (p_4 p_6), (p_6 p_3)$, and $(p_3 p_1)$, as depicted by solid line. However, we can take another set of edges that surround this loop: $(p_1 p_2), (p_2 p_4), (p_4 p_6), (p_6 p_5), (p_5 p_3)$, and $(p_3 p_1)$, depicted as dashed lines. To avoid the ambiguity when defining the set of edges that surround the loop, the volume optimal cycle is defined as the loop that has minimum number of triangles inside. In our example, the first loop has  5  edges, and can be divided into 3 triangles: $(p_1 p_2 p_4)$, $(p_1 p_4 p_6)$, and $(p_1 p_6 p_3)$. Of course, there are many other ways to decompose this loop into triangles, however, the number of triangles is uniquely determined. The second loop consists of 6 edges, and 4 triangles are needed to construct this loop. Therefore, we choose the first loop structure as the "volume optimal cycle."

We frequently designate loops that emerge in PH as "generators". From one atomic conformation, we obtain several generators. In PH, the generator birth and death distributions give important insights into the dataset structure. To visualize this distribution, we use the scatter plot of births and deaths, which is called a persistence diagram. Another visualization method is persistent barcodes, in which horizontal lines represent generator births and deaths. We will present several barcode plots in Sec.\ref{Sec:Result}.  

As we have mentioned in Sec.~\ref{Sec:Introduction}, the number of generators strongly depends on the atom configuration. Even if the number of atoms is the same, the number of generators can differ. This fact makes it difficult to combine machine-learning techniques with PH. We attempted to overcome this challenge by introducing a "TFV" composed of the volume optimal cycles (see below). The calculation of births, deaths, and volume optimal cycles was performed by HomCloud ver.1.2.1\cite{Homcloud}.

PH is strongly related to Betti numbers, which are topological invariants in mathematics. In topology, the $k$-th Betti number is defined as the rank of $k$-th homology groups. In our case, $k=1$, it is the number of "loops." Therefore, if we put balls with radius $r$ at $p_1, p_2, \cdots, p_n$, the first Betti number of this set is the number of generators whose births and deathes are smaller and larger than $r$, respectively. 

Before concluding this subsection, we discuss the use of higher-degree PH. In homology, "degree" is the dimension of "boundaries," and the PH with degree 2 is used to investigate the vacancies surrounded by triangles. Degree 2 PH often plays an important role in material science as it provides information on the voids. However, Xia and Wei found that PH with degree 2 gives little information on protein structure\cite{Xia2014}. Further, they revealed that both $\alpha$-helices and $\beta$-sheets give provide voids when analyzing  C$_\alpha$ atoms as a point cloud. The native chignolin structure contains only one $\beta$-sheet and no tertiary structure. Thus, we choose to ignore higher-degree PH in this study.

\subsection{Construction of topological feature vector (TFV)}

\begin{figure}
    \centering
    \includegraphics[width = \textwidth]{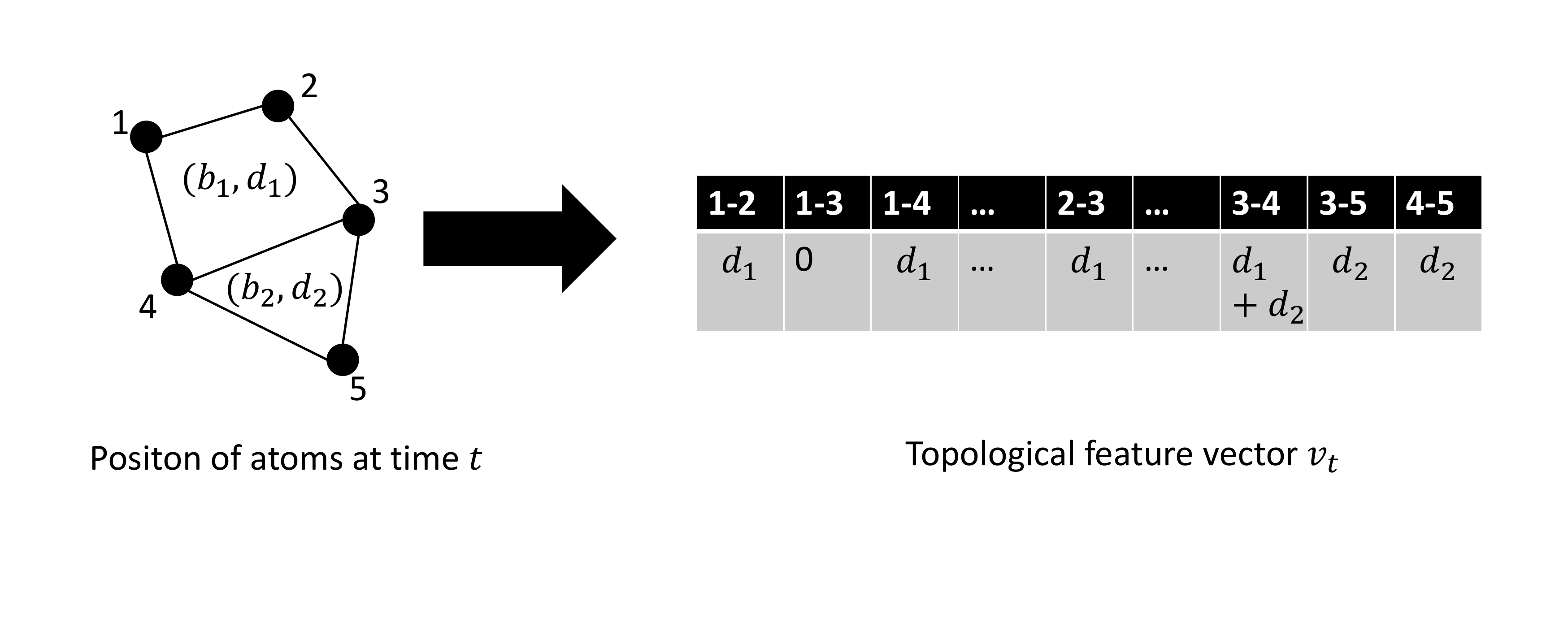}
    \caption{Schematic description of the TFV.}
    \label{fig:topological_feature}
\end{figure}

Using the loop information, we defined a "TFV," $\bm v$, which describes the point cloud topology. First, for each edge $E=(p_i p_j)$, we listed the generators $g_k$, whose volume optimal cycles include $E$. We then calculated the "importance" of the edge $E$ as the sum of the deaths of $g_k$. If an edge was not included in any volume optimal cycles, we set the edge "importance" as 0 (Fig.\ref{fig:topological_feature}). By this method, we obtained the TFV, whose dimension is $M = n(n-1)/2$.

Feature vector construction was similar to a "bag-of-words" and "term frequency-inverse document frequency," which are standard methods used in natural language processing\cite{manning2010introduction}. These methods regard a document as a multiset of terms and calculate the "importance" for each term. In our approach, edges were defined as the terms that describe the protein shape.

There are several possible methods to construct a TFV from the volume optimal cycles. For example, we could create another TFV using births instead of deaths. Lifetime, the difference between death and birth, is also often used as an important PH variable. In this study, we examined the results of birth-based and death-based TFV. These qualitatively yielded the same result. Alternatively, we could use the products of deaths instead of sums of deaths. In this study, we used the sum of deaths for simplicity. Indeed, there may be more a complex and sophisticated definition of the TFV (See Sec.~\ref{Sec:Conclusion}).

\subsection{Dimension reduction by non-negative matrix factorization}

The dimensions of a TFV are generally high, making dimension reduction using PCA or another method useful. We primarily employed NMF to reduce the dimensionality of TFV\cite{Lee1999}. We assumed that the $M$-dimensional TFV at time $t=t_1, t_2, \cdots t_N$ are $\bm{v}_1, \bm{v}_2, \cdots, \bm{v}_N$, where $\bm{v}_i = (v_{1i},v_{2i},\cdots,v_{Mi})^t$, where $(\cdots)^t$ represents the transverse of the matrix, respectively. In NMF, we attempted to reduce the system into $L$-dimensional space, under the assumption that both coefficients and bases are non-negative. We calculated $M \times L$ non-negative matrix $\bm W=(w_{ij})$ and a $L\times N$ non-negative matrix $\bm H = (h_{ij})$ that minimized $||\bm V-\bm W \bm H||$, where $||\cdots ||$ represents the Frobenius norm. Using this method, we can approximated $\bm{v}_i \approx \sum_{k=1}^{L} \bm{w_k} h_{ki}$, where $\bm{w}_k = (w_{1k}, w_{k2},\cdots, w_{Mk})^t$ are the bases of the reduced space.

Compared with PCA, NMF has several advantages. First, when we reconstructed TFVs from the information in reduced spaces, NMF consistently  generated non-negative vectors. Both NMF and PCA attempt to approximate the feature vector $\bm{v}$ by the linear combination of several bases vectors $\bm{e}_i$: $\bm{v} \approx \sum_i c_i \bm{e}_i$. In NMF, we set $c_i \ge 0$ and $\bm{e}_i$ to be non-negative, and approximated $\bm{v}$ was also non-negative. Conversely, certain components in $\sum_i c_i \bm{e}_i$ can be negative in PCA. When $\bm{v}$ is defined as non-negative vectors, understanding large negative components in $\sum_i c_i \bm{e}_i$ is difficult.
Another advantage of NMF is that the bases can capture important local features. Though there is no theoretical explanation, the application of NMF to face-recognition problems shows that NMF can extract localized characteristics such as noses or eyes, while PCA captures non-local structures \cite{Lee1999}. In the case of protein folding analysis, the ability of NMF to capture local structure is desirable. For example, we considered the folding of proteins with several secondary $\alpha$ helices and $\beta$ sheet structures. In this case, it is natural to assume that the secondary structure formation does not occur simultaneously. In NMF, we expected several bases to represent secondary structures. However, if we used PCA for decomposition, each basis represents the complex structural change, such as the disappearance of several helices and appearance of several sheets. Therefore, we need further investigation to understand the formation of these structures. 

Though useful, NMF has several disadvantages. First, the NMF decomposition is not unique. Suppose that $\bm{W}$ and $\bm{H}$ are non-negative matrices. If both $\bm{A}$ and $\bm{A}^{-1}$ are non-negative matrices, then $\bm{W}^\prime = \bm{WA}$ and $\bm{H}^\prime = \bm {A^{-1} H}$ are non-negative, and we obtain another decomposition $\bm{V}\approx \bm{W^\prime H^\prime}$. In practice, when the feature matrix is sparse and we  initialize  $\bm{W}$ and $\bm{H}$ by a non-negative double singular value decomposition, then optimization with a coordinate descent solver generally yields small residue $||\bm{V}-\bm{WH}||$ with low computational costs\cite{Boutsidis2008}. This method is deterministic and free from the problem caused by non-uniqueness of NMF decomposition. Since our feature vector is sparse, we applied this initialization and optimization method.     
NMF also has the ambiguity of "scales." We  can "rescale" the basis $\bm{w_i}$; $\bm{w}^\prime_i = \alpha_i \bm{w}_i$ and $h^\prime_{ij}=\alpha^{-1}_i h_{ij}$ where $\alpha_i$'s are positive constants,  which provides another decomposition. Here, we scaled $\bm{w}$s so that $||w=1||$.
Thus, $\bm{w}$ can be assumed as a dimensionless vector, and $h_{ij}$ has the same dimension as births and deaths.

Another disadvantage of NMF is the need to determine the rank of reduced space $L$ {\it a priori}. Though there is no de facto standard to estimate rank $L$, several methods are proposed \cite{Brunet2004, Hutchins2008}. In our study, we used the method proposed by Hutchins {\it et al.}\cite{Hutchins2008} who showed that if the dataset is random, the residual sums of squares (RSS) between $\bm{V}$ and $\bm{WH}$  decreases linearly with rank $r$, and proposed to use $L$ at the inflection point. We performed these calculations using scikit-learn 0.19.1 and NMF 0.21.0\cite{scikit-learn, Gaujoux2010}.

\subsection{Chignolin molecular dynamics simulation}

Using the method described in Mitsutake and Takano\cite{doi:10.1063/1.4931813}, we conducted MD simulations of aqueous chignolin near a transition temperature. We placed one chignolin molecule, 2 Na$^{+}$ atoms, and 3674 H$_2$O molecules in a cube and set the temperature and pressure at 450K and 1 atm, respectively. After energy minimization and equilibration for 50--ns, we performed a 1$\mu$s NPT-constant MD simulation. We captured snapshots of the molecules every 10 ps to create 100,000 samples. In this simulation, we used the ff99SB force field and TIP3P models for the water molecules. From each snapshot, we obtained the coordinates of ten C$_\alpha$ atoms in chignolin and performed the PH analysis. The simulation was conducted using GROMACS 16.4\cite{Berendsen1995}.

\section{Result}\label{Sec:Result}

In this study, we performed PH analysis of a point cloud composed of ten C$_{\alpha}$-atoms.
We calculated the TFVs from snapshots of  chignolin and reduced the configuration into low dimensional spaces by PCA and NMF.

\subsection{Analysis using TFV}

\begin{figure}
    \centering
    \subfigure[]{\includegraphics[width=.45\textwidth]{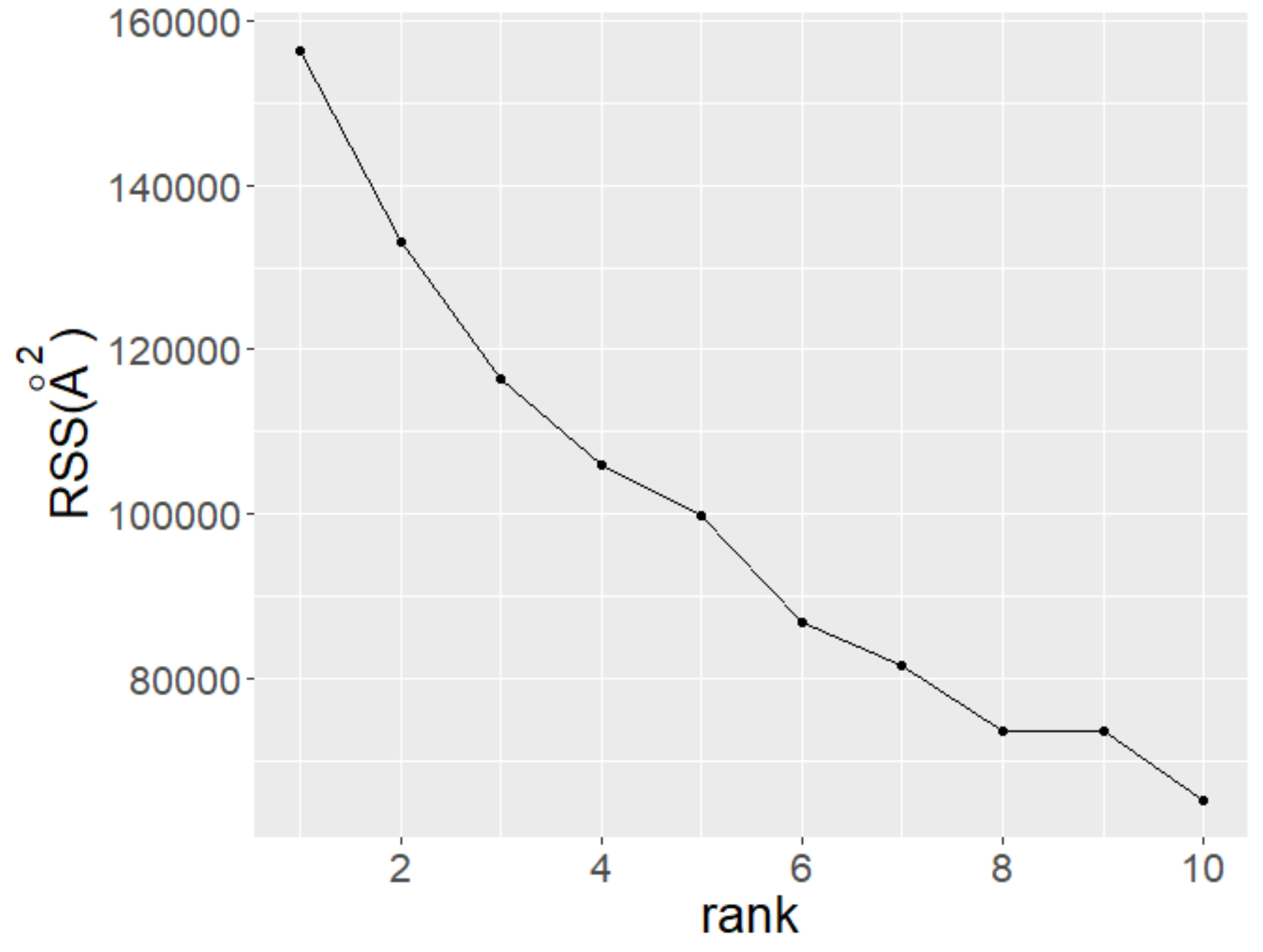}}
    \subfigure[]{\includegraphics[width =.45\textwidth]{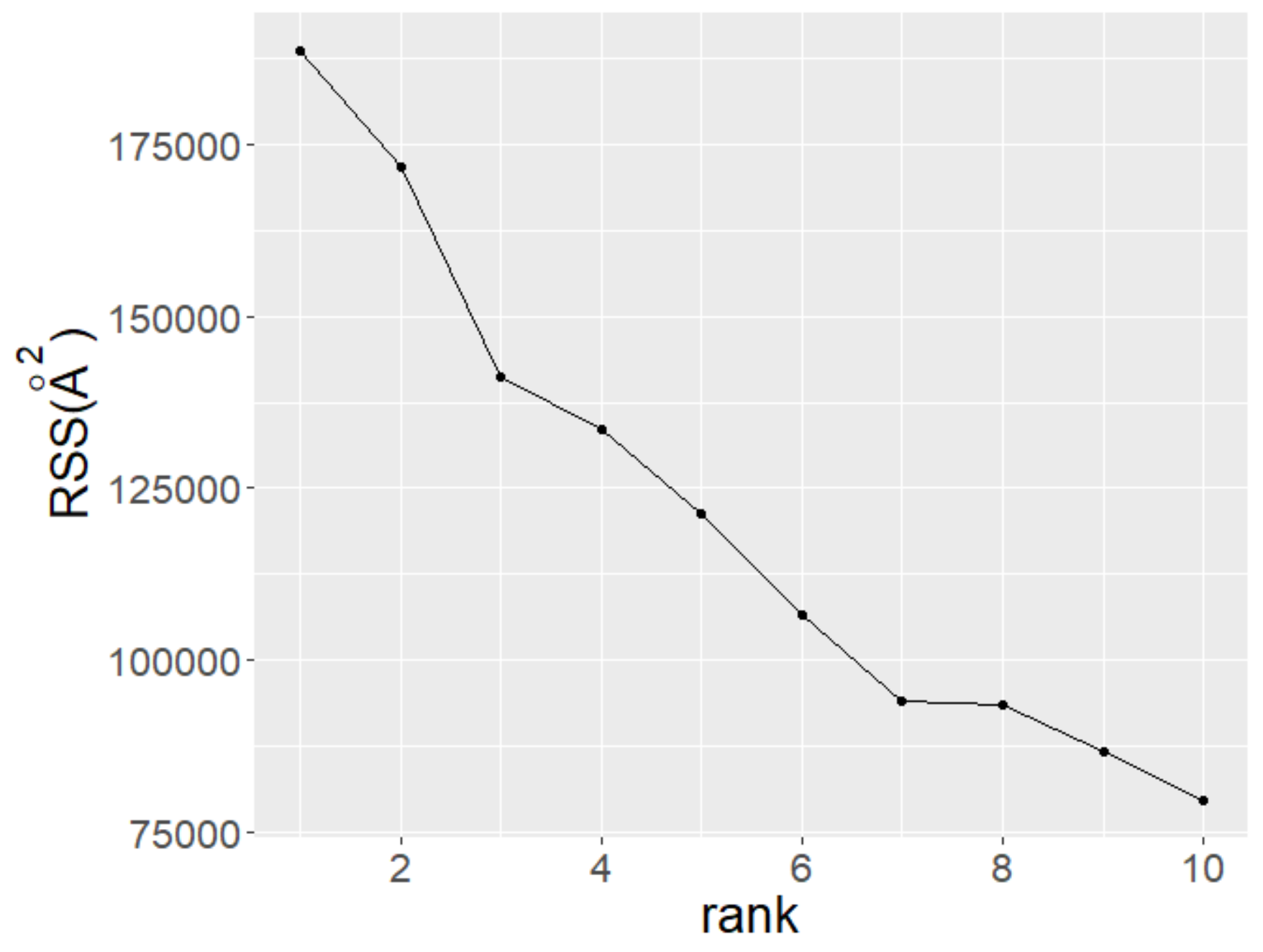}}
    \caption{Plot of residual sums of squares for rank $L = 1, 2, \cdots 10$: (a) TFV is constructed from births. (b) TFV is constructed from deaths.}
    \label{fig:rss}
\end{figure}

To carry out NMF analysis, we first determined the rank of reduced space. To determine the rank, we calculated RSS to determine the rank of reduced space $L$. To reduce the computational cost, we randomly selected 1,000 samples from our dataset, and carried out NMF for $L= 1, ,2 \cdots, 10$. The obtained RSS is shown in Fig.~\ref{fig:rss}. In Fig.~\ref{fig:rss}(a), when we used births to construct the TFV, the RSS rapidly decreased as $L$ increased from 1 to 3, and slowly decreased for $L>3$. This result indicates that $L=2$ or 3 are the best reduced space ranks. When we used deaths to construct the TFV, the RSS shown in Fig.~\ref{fig:rss}(b) was obtained, again suggesting that $L=2$ or 3 is the best rank of reduced space.

\begin{figure}
    \centering
    \subfigure[birth-based TFV, rank = 2]{\includegraphics[width= .45\textwidth]{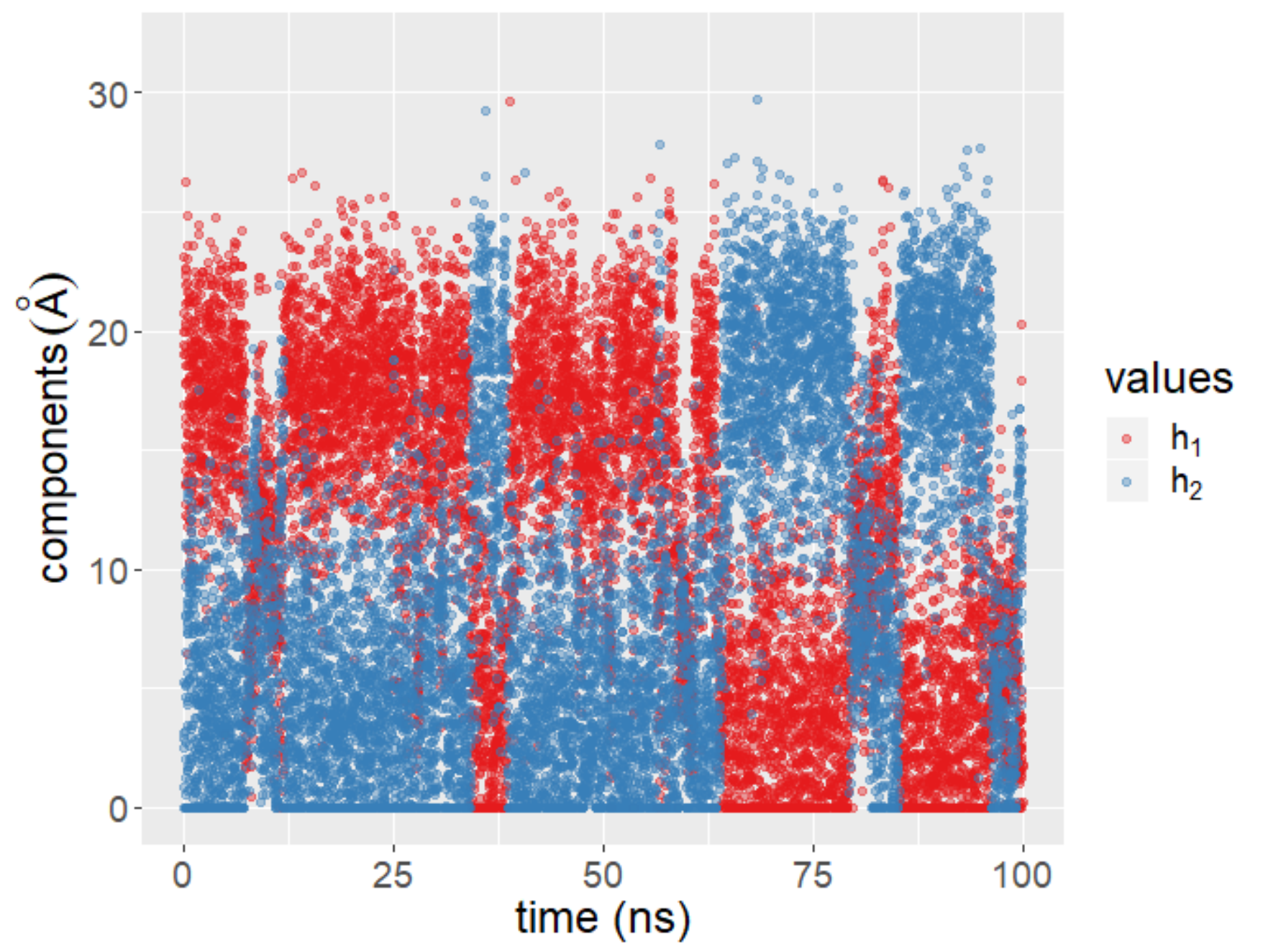}\label{time-series_birth2}}
    \subfigure[death-based TFV, rank = 2]{\includegraphics[width =.45\textwidth]{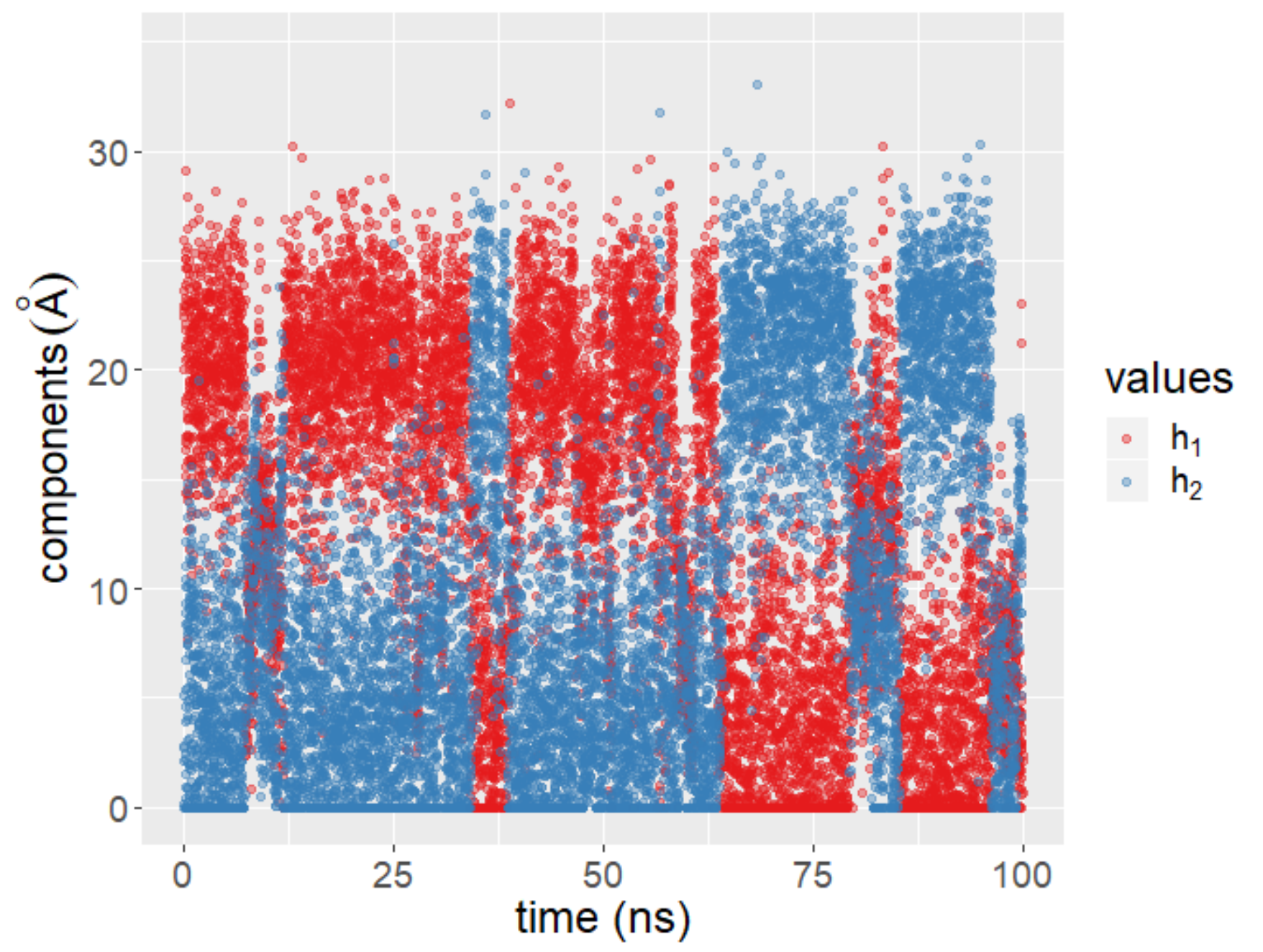}\label{time-series_death2}}
    \subfigure[birth-based TFV, rank =3]{\includegraphics[width= .45\textwidth]{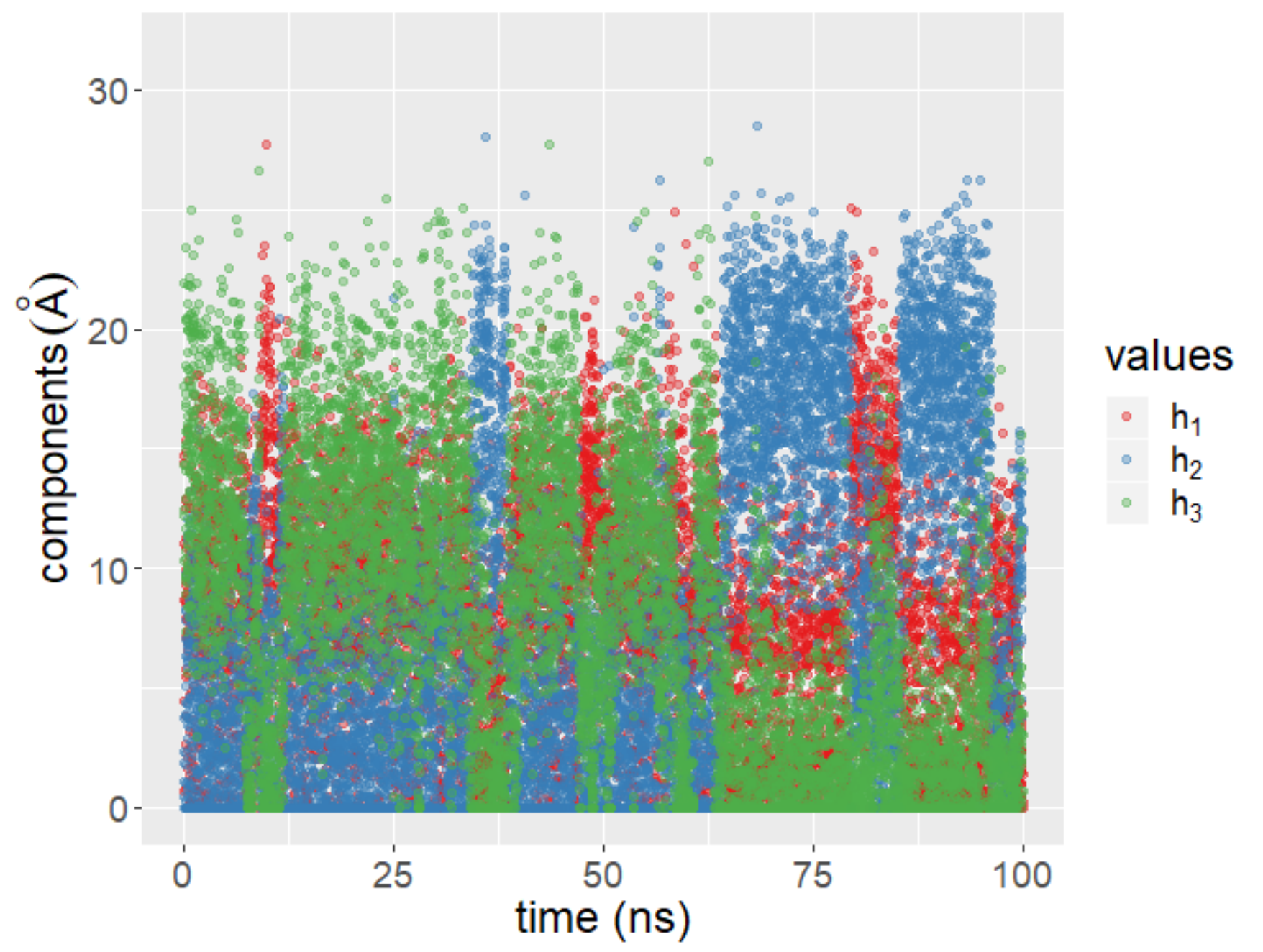}\label{time-series_birth3}}
    \subfigure[death-based TFV, rank=3]{\includegraphics[width =.45\textwidth]{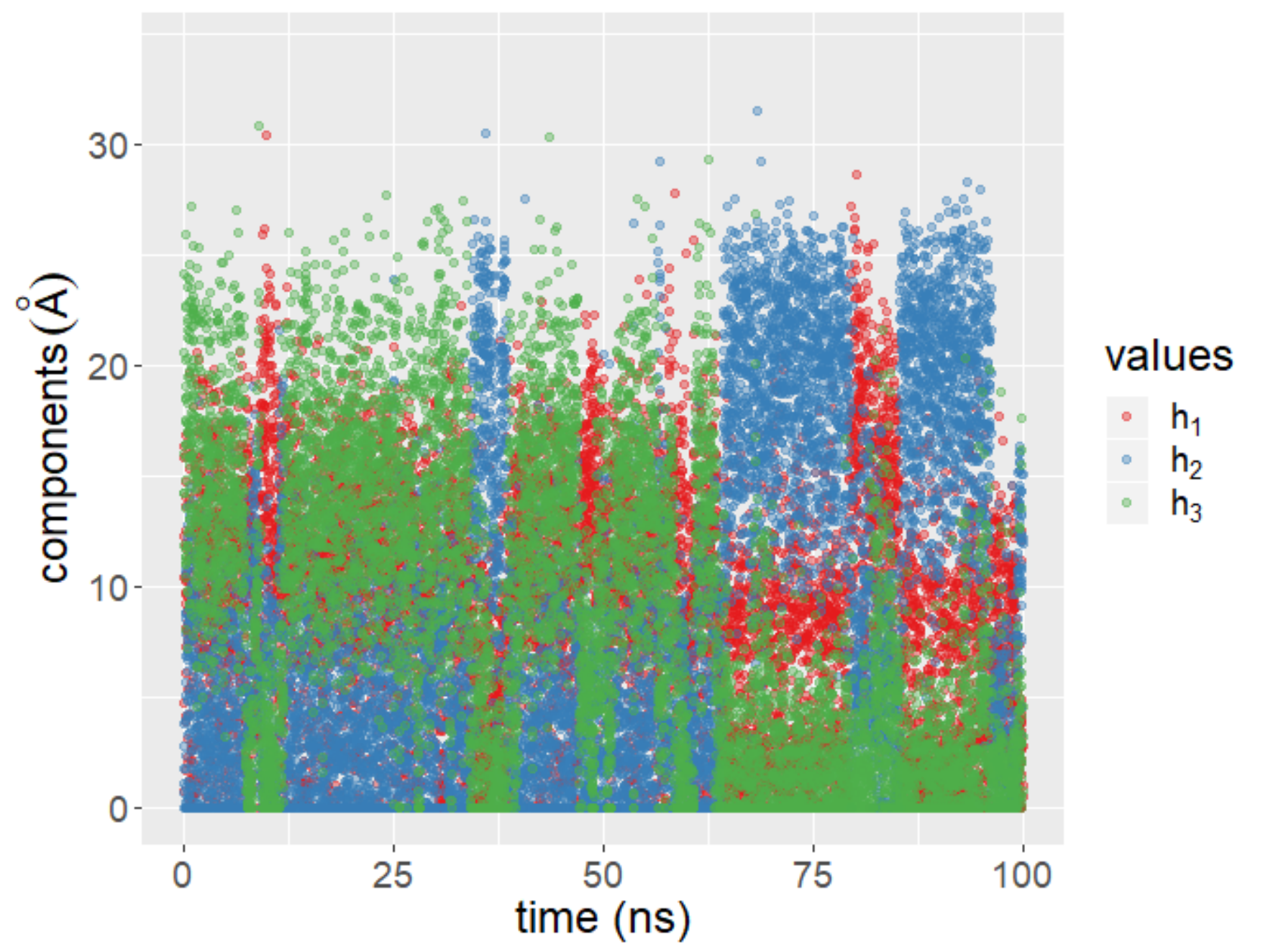}\label{time-series_death3}}
    \subfigure[birth-based TFV, rank = 4]{\includegraphics[width= .45\textwidth]{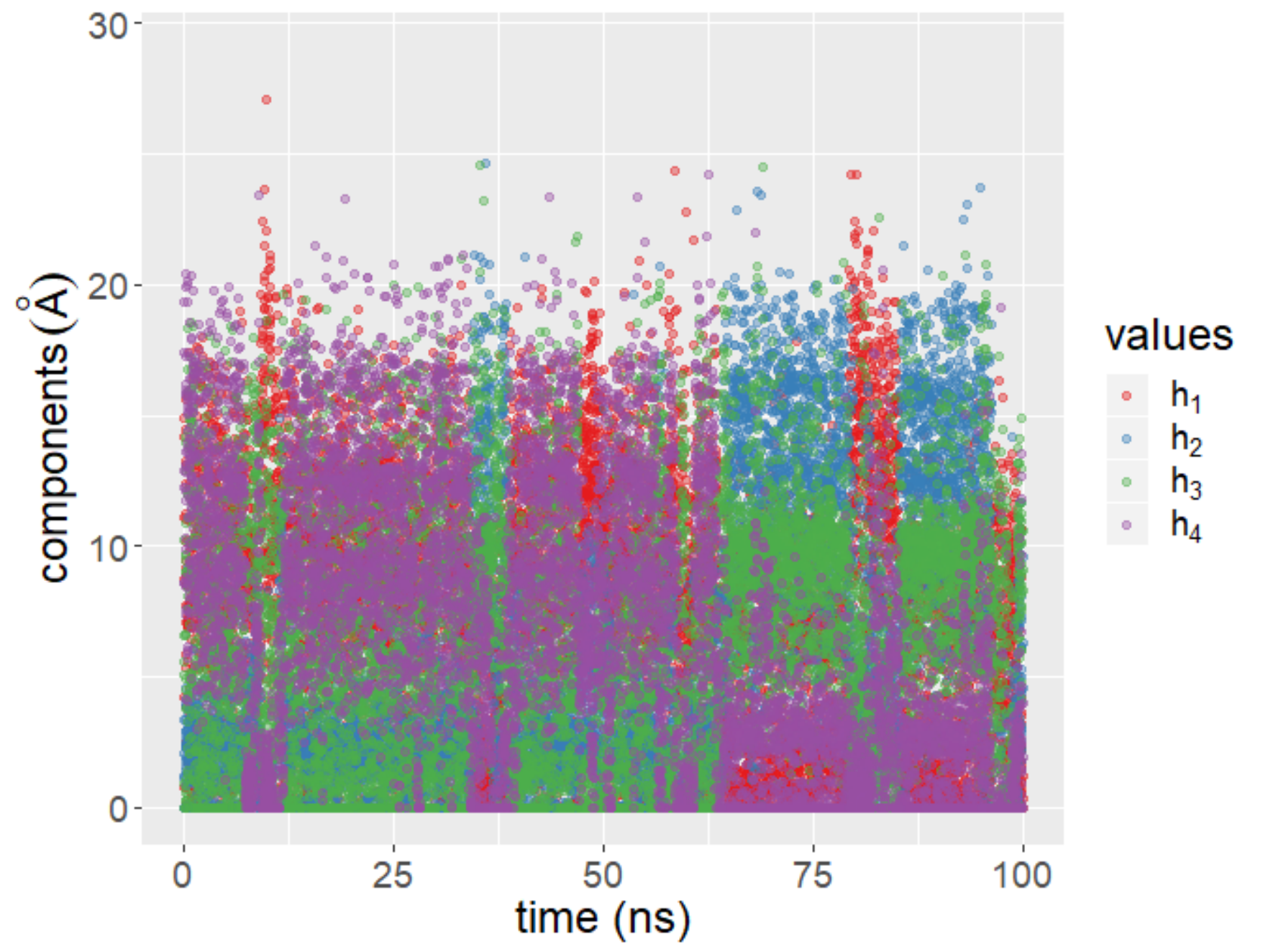}\label{time-series_birth4}}
    \subfigure[death-based TFV, rank = 4]{\includegraphics[width =.45\textwidth]{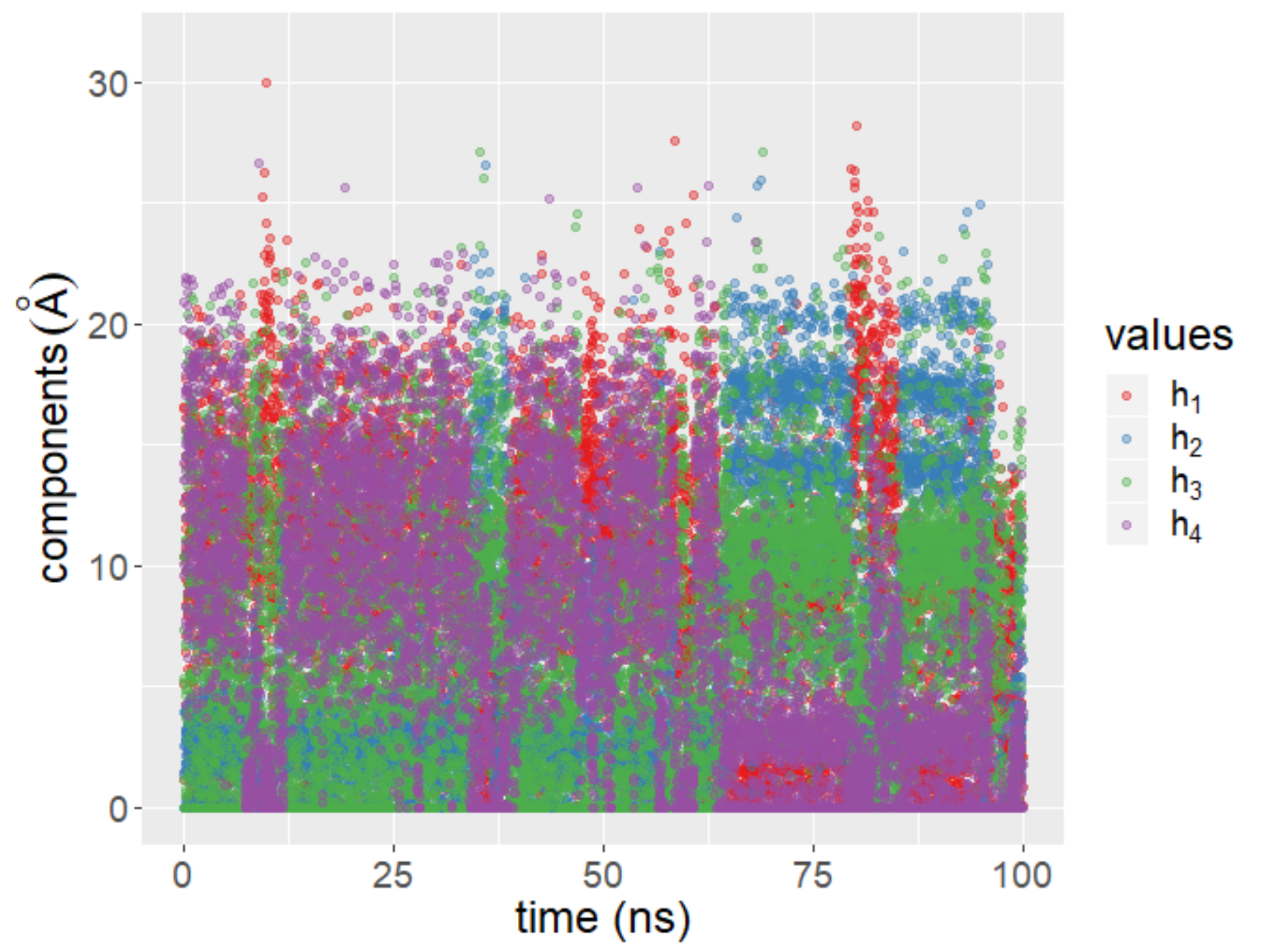}\label{time-series_death4}}
    \caption{Dynamics of the TFV in  reduced spaces. (a) rank=2, birth-based TFV. (b) rank=2, death-based TFV, (c) rank=3, birth-based TFV, (d) rank=3, death-based TFV, (e) rank=4, birth-based TV, (f) rank =4, death-based TFV.}
    \label{fig:time-series}
\end{figure}

To investigate the effect caused by changing $L$, we compared the results obtained by TFVs constructed from births and deaths for $L= 2$, 3 and 4. Fig.~\ref{fig:time-series} represents the dynamics in the reduced space for $t=0$ to 100ns. In this figure, $h_k$ at time $t$ represents the value of $h_{ki}$, where $i$ is the TFV index obtained from the snapshot at time $t$. Fig.~\ref{time-series_birth2} shows the dynamics when $L=2$ and births were used to construct the TFV. Clearly, there are two phases: the first shows that $10 \text{\AA}  \lesssim h_1 \lesssim 30 \text{\AA}$ while $h_2 \lesssim 10$ \AA; the other shows that $10 \text{\AA} \lesssim h_2 \lesssim  30$ \AA while $h_1 \lesssim 10$ \AA. We also noted that short periods occur where both $5 \text{\AA} \lesssim h_1, h_2 \lesssim 15 \text{\AA}$. Therefore, it seems that there are two or three phases.
This result is not modified when deaths are used instead of births to define TFV, as shown in Fig. ~\ref{time-series_death2}. The correlation between $h_1$ in Fig. ~\ref{fig:time-series}(a) and (b) was 0.9967, and the correlation between $h_2$ was 0.9968.
 Fig.~\ref{time-series_birth3} and \ref{time-series_death3} show the plots when $L=3$. When we compared Fig.~\ref{time-series_birth2} and ~\ref{time-series_birth3}, we found that at the second phase in Fig.\ref{time-series_birth2} where $h_1 \lesssim 10 \text{\AA}$ and $10 \text{\AA} \lesssim h_2 \lesssim 30 \text{\AA}$,  $h_2$ in Fig.~\ref{time-series_birth3} is large: $ 10 \text{\AA} \lesssim h_2 < 30 \text{\AA}$ while $h_1, h_3 \lesssim 10 \text{\AA}$. Additionally, when $10 \text{\AA} \lesssim h_1 \lesssim 30 \text{\AA}$ in Fig.~\ref{time-series_birth2}, $h_2\lesssim 10 \text{\AA}$ in Fig.~\ref{time-series_birth3}, while no clear difference was observed  between $h_1$ and $h_3$. Third, when $5 \text{\AA} \lesssim h_1, h_2 \lesssim 15 \text{\AA}$ in Fig.~\ref{time-series_birth2}, $10 \text{\AA} \lesssim h_1\lesssim 30 \text{\AA}$ while $h_2, h_3\lesssim 10 \text{\AA}$ in Fig.~\ref{time-series_birth3}. However, the third observation seems controversial, due to the short phase duration. From these observations, we found no additional stable phase by increasing the rank $L$ from 2 to 3. However, the analysis for $L=3$ may be useful to understand the details of the phase where both $h_1$ and $h_2$ are small in Fig.\ref{time-series_birth2}. These observations are not modified when we set rank $L=4$, shown in Fig.~\ref{time-series_birth4} and ~\ref{time-series_death4}. Thus, we used rank $L=2$ and deaths to construct TFVs.

\begin{figure}
    \centering
    \subfigure[Density of NMF-reduced states]{\includegraphics[width=.45\textwidth]{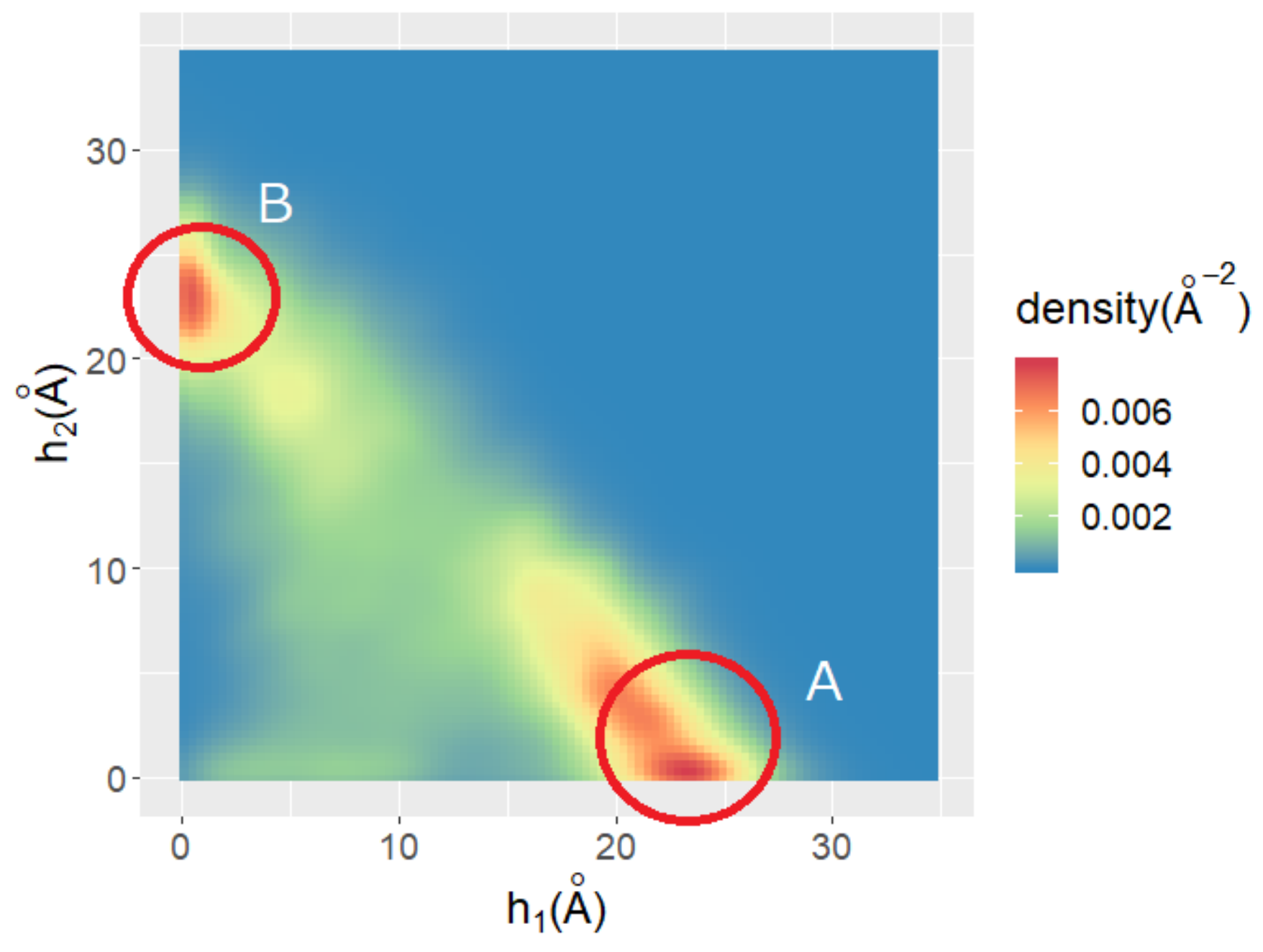}\label{density_nmf}}
    \subfigure[Density of PCA-reduced states]{\includegraphics[width =.45\textwidth]{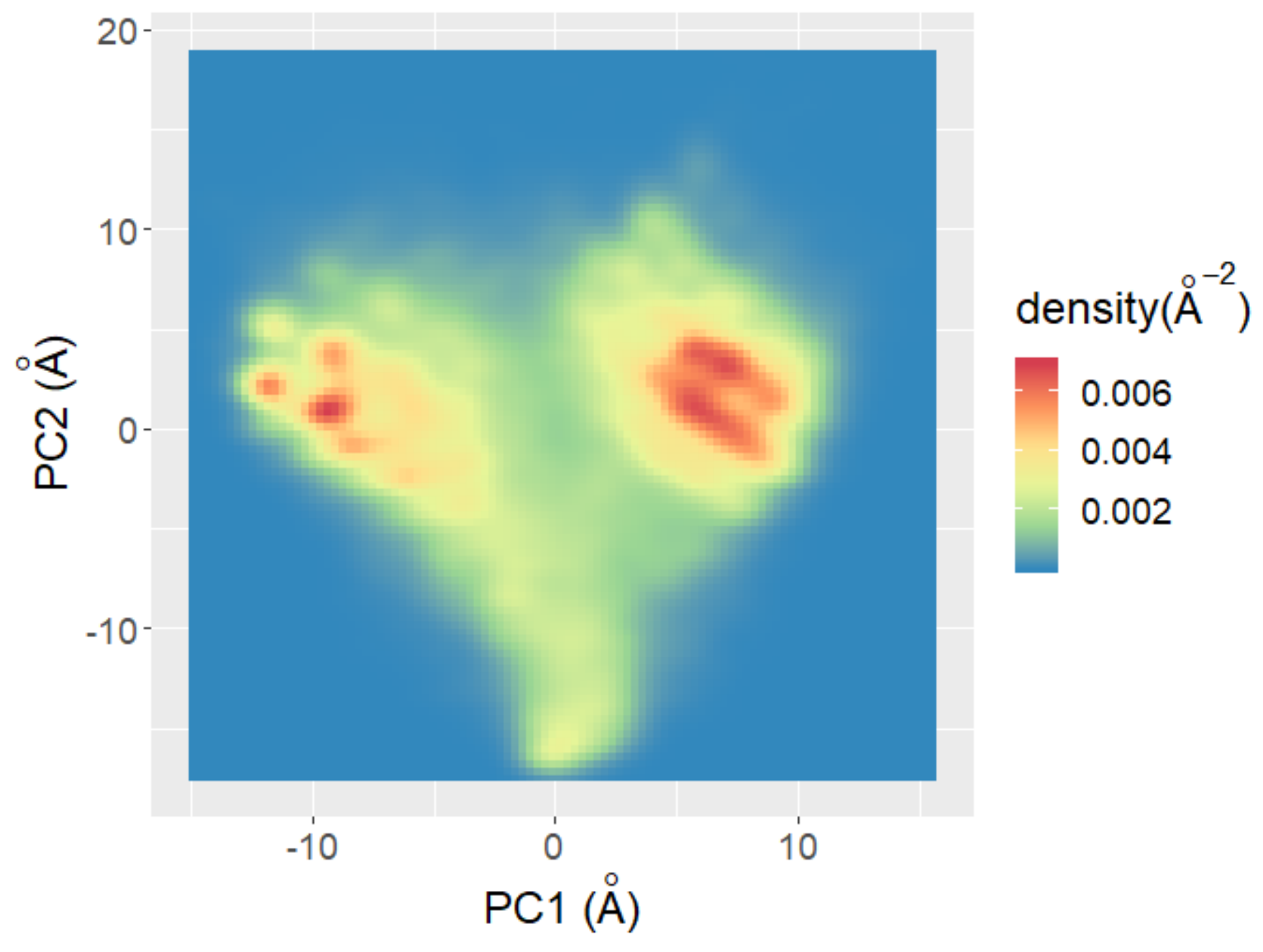}\label{density_pca}}
    \caption{Density of states in reduced spaces. (a) NMF reduction. (b) PCA reduction.}
    \label{fig:density}
\end{figure}

Fig.~\ref{density_nmf} shows the TFV distribution in the reduced space by NMF, $L=2$. We observed two large peaks at $(h_1, h_2) \sim (25 \text{\AA} ,0 \text{\AA})$ and $(0 \text{\AA},25 \text{\AA})$, which we designated, region A and B, respectively. The density was high along the straight line connecting these peaks, with the minimum at $(h_1, h_2) \sim (10 \text{\AA},15 \text{\AA})$. We also noted that the density  in the area around $(h_1, h_2)\sim (5 \text{\AA},0 \text{\AA})$ was high.
This result is consistent with the results obtained by PCA of TFVs shown in Fig.~\ref{density_pca}. Here, we clearly identified two clusters, which are distinguished by 1st principal components. The correlation between 1st principle component and $h_1, h_2$ are 0.941 and -0.927 respectively. Therefore, the right and left clusters in Fig.~\ref{density_pca} correspond to clusters at region A and B in Fig.~\ref{density_nmf}, respectively. Though NMF and PCA gave qualitatively similar results, we only discuss the results of NMF in the following sections of this paper as they more clearly reveal protein structures; whereaas PCA indicated only the "difference" of these two clusters. As described in the Sec.\ref{Sec:Method}, we found negative components when we reconstructed the TFV from PCA, which exacerbates the difficulty in understanding the protein structure.
The NMF result in Fig.\ref{density_nmf} indicates that the peaks of clusters are nearly on the $h_1$ and $h_2$ axes, respectively. Thus, we can infer the "typical" chignolin TFVs in each cluster by checking $\bm{w}_1$ and $\bm{w}_2$ directly.

\begin{figure}
    \centering
    \includegraphics[width=\textwidth]{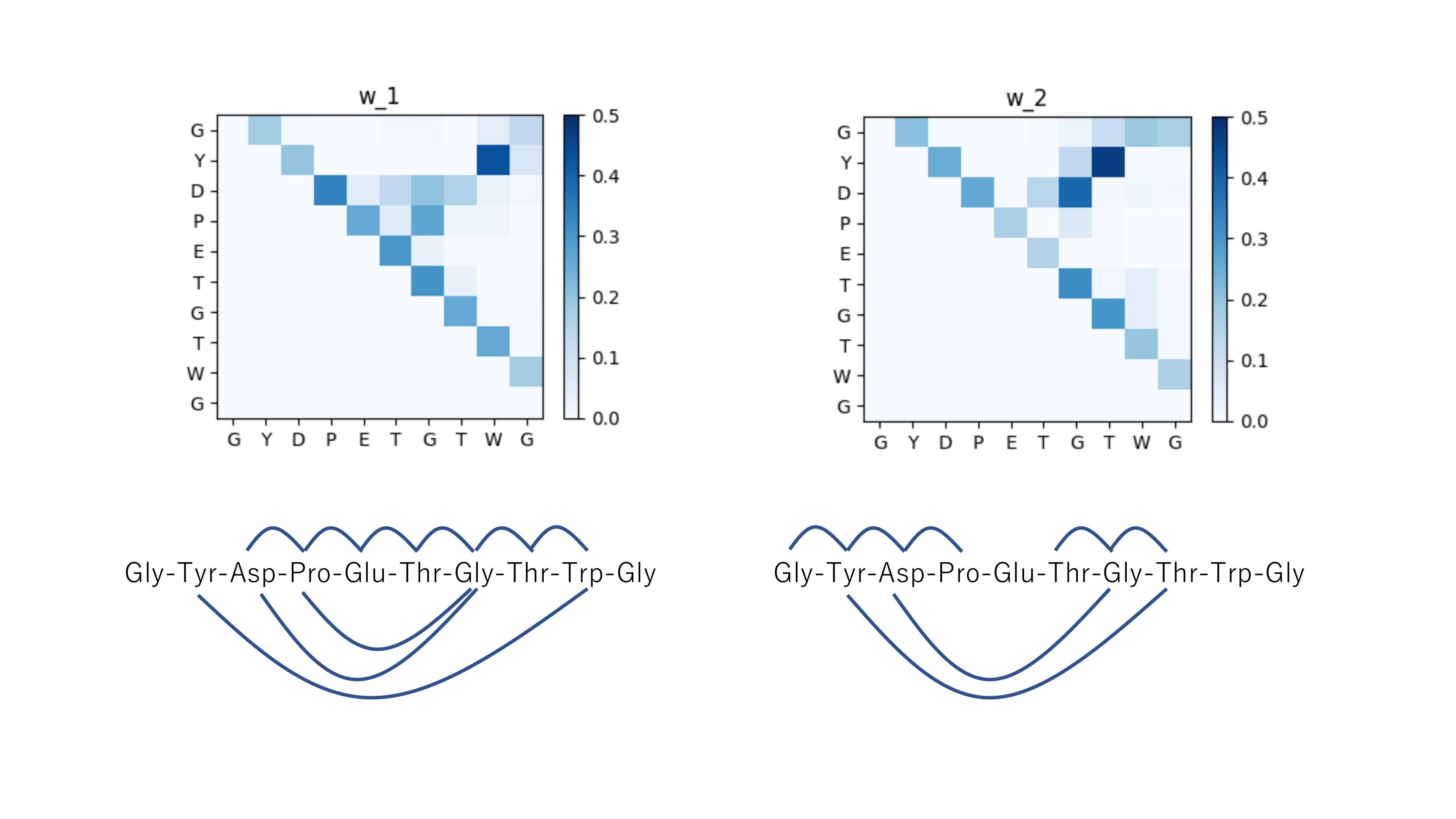}
    \caption{Components of $\bm{w}$  for rank =2. We also represent the edges whose corresponding components are larger than 0.2 by blue lines. }
    \label{fig:nmf_components}
\end{figure}

To investigate the structures at the two density peaks, we examined the bases $\bm{w}_1$ and $\bm{w}_2$ obtained by NMF, as shown in Fig.~\ref{fig:nmf_components}. Chignolin has ten C$_\alpha$ atoms and the TFV dimension is 45, which is the number of C$_\alpha$ atom pairs. In the top panel of this figure, we plotted the value of each component in $\bm{w}_1$ and $\bm{w}_2$. Darker squares represent that the corresponding edge component is large. We omitted the component in the lower left triangle. For example, in the case of $\bm{w}_1$, the color at the second row-ninth column is dark, which implies that the component of $\bm{w}_1$ that corresponds to the Tyr2--Trp9 pair is large. From this figure, when the protein was in region A in Fig.\ref{density_nmf}, the edge between Tyr2 and Trp9 made a large contribution to loop formation, since the corresponding feature vector was well approximated by $h_1 \bm{w}_1$. In the bottom panel of this figure, edges with corresponding components in $\bm{w}$ larger than 0.2 are indicated by blue lines. We note that $\bm{w}_i$'s are dimensionless, as discussed in Sec.~\ref{Sec:Method}.  

From this figure, we noted that the adjacent amino acid pairs, such as Gly1--Tyr2 or Asp3--Pro4, made a large contribution to the cycle formation. This is natural since the distances between adjacent amino acids are short, due to chemical bonding. There was a large difference between $\bm{w}_1$ and $\bm{w}_2$ on the components corresponding to Tyr2--Trp9, Tyr2--Thr8, and Pro4--Gly7.
In particular, the components corresponding to Tyr2--Trp9 and Tyr2--Thr8 showed clear differences;  $\bm{w}_1(\mbox{Tyr2--Trp9}) = 0.43 $, while that of $\bm{w}_2(\mbox{Tyr2--Trp9}) = 0$. $\bm{w}_1(\mbox{Tyr2--Thr8})=0$ while $\bm{w}_2(\mbox{Tyr2--Thr8}) = 0.47$.
Therefore, $h_1$ becomes large if there is a loop whose volume optimal cycle includes Tyr2--Trp9, and $h_2$ becomes large when a loop forms whose volume optimal cycle includes Tyr2--Thr8. These results are depicted in the bottom panel of this figure. To determine which state corresponds to the native structure, prior knowledge on the native state is needed. When we applied PH to the native structure, we found that Tyr2--Trp9 was dominant in the native state. The TFV component for Tyr2--Trp9 was 1.249, while that for Tyr2--Thr8 was 0.130. Therefore, we conclude that the cluster at region A in Fig.~\ref{density_nmf} corresponds to the native state, whereas the cluster at region B corresponds to the misfolded state. Concerning the state around $(h_1, h_2) \sim (5.0 \text{\AA}, 0 \text{\AA})$, we noted that a small $h_1$ and $h_2$ implies that there are no loops. Since the  feature vectors $\bm v$ were approximated as  $ \bm{v} \sim \sum_i h_i \bm w_i$, $h_1 = h_2 = 0$ suggests that $\bm{v} \sim \bm{0}$. Therefore, we hypothesized that the state with a small $(h_1, h_2)$ is unfolded.

\begin{figure}
    \centering
    \subfigure[t=160ns]{\includegraphics[width=0.3\textwidth]{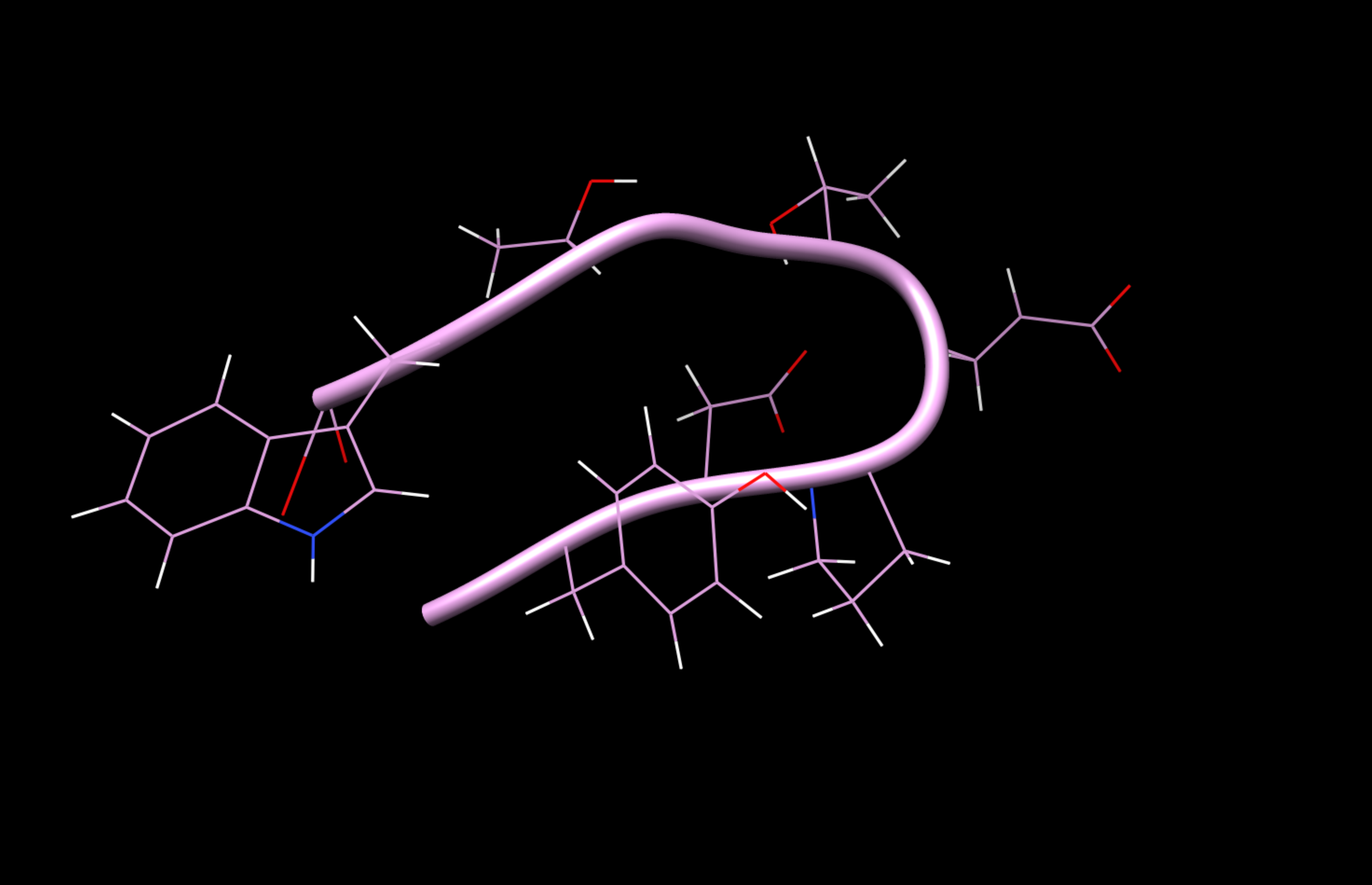}\label{sample_native}}
    \subfigure[t=120ns]{\includegraphics[width=0.3\textwidth]{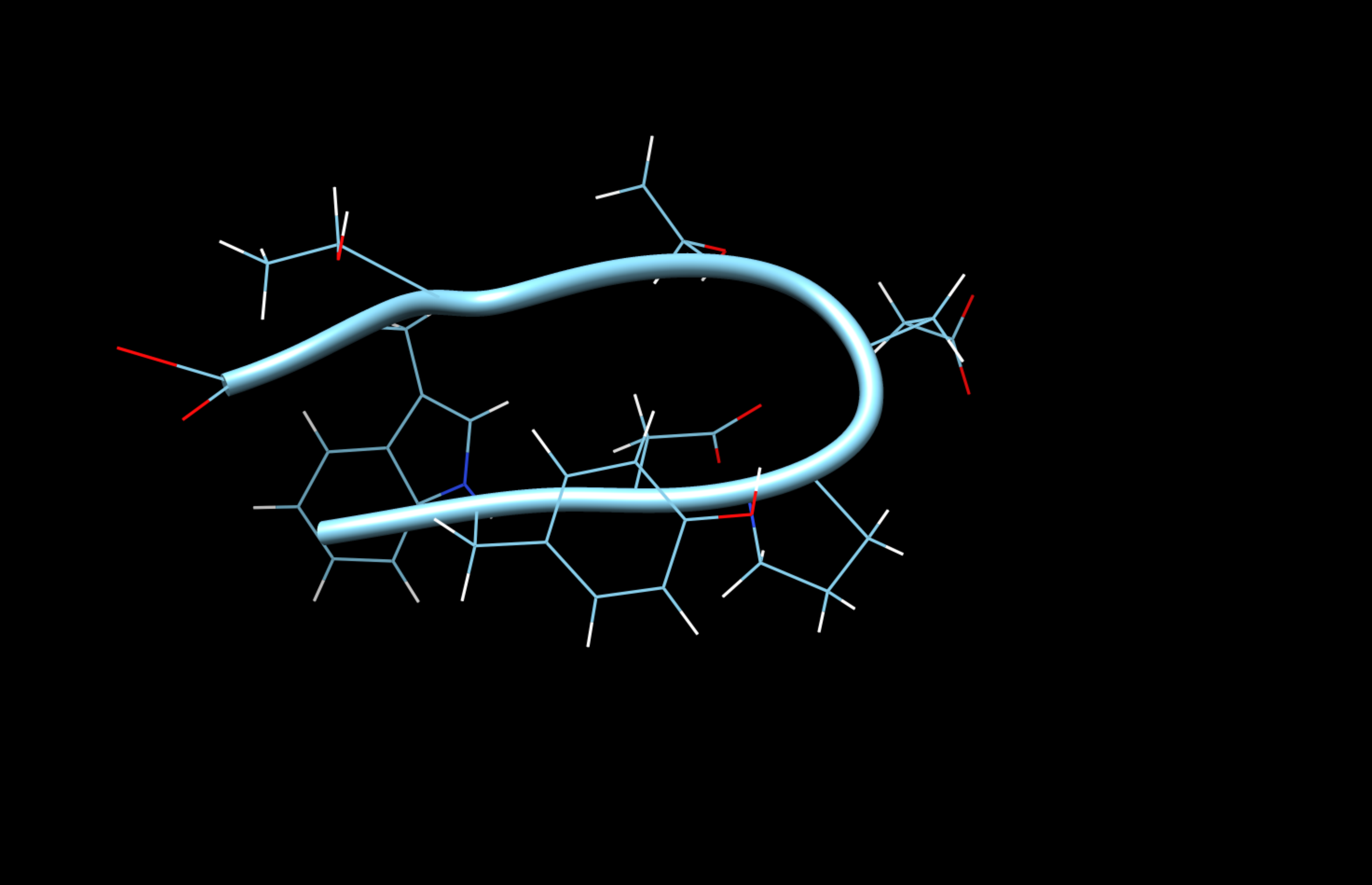}\label{sample_misfold}}
    \subfigure[t=60ns]{\includegraphics[width=0.3\textwidth]{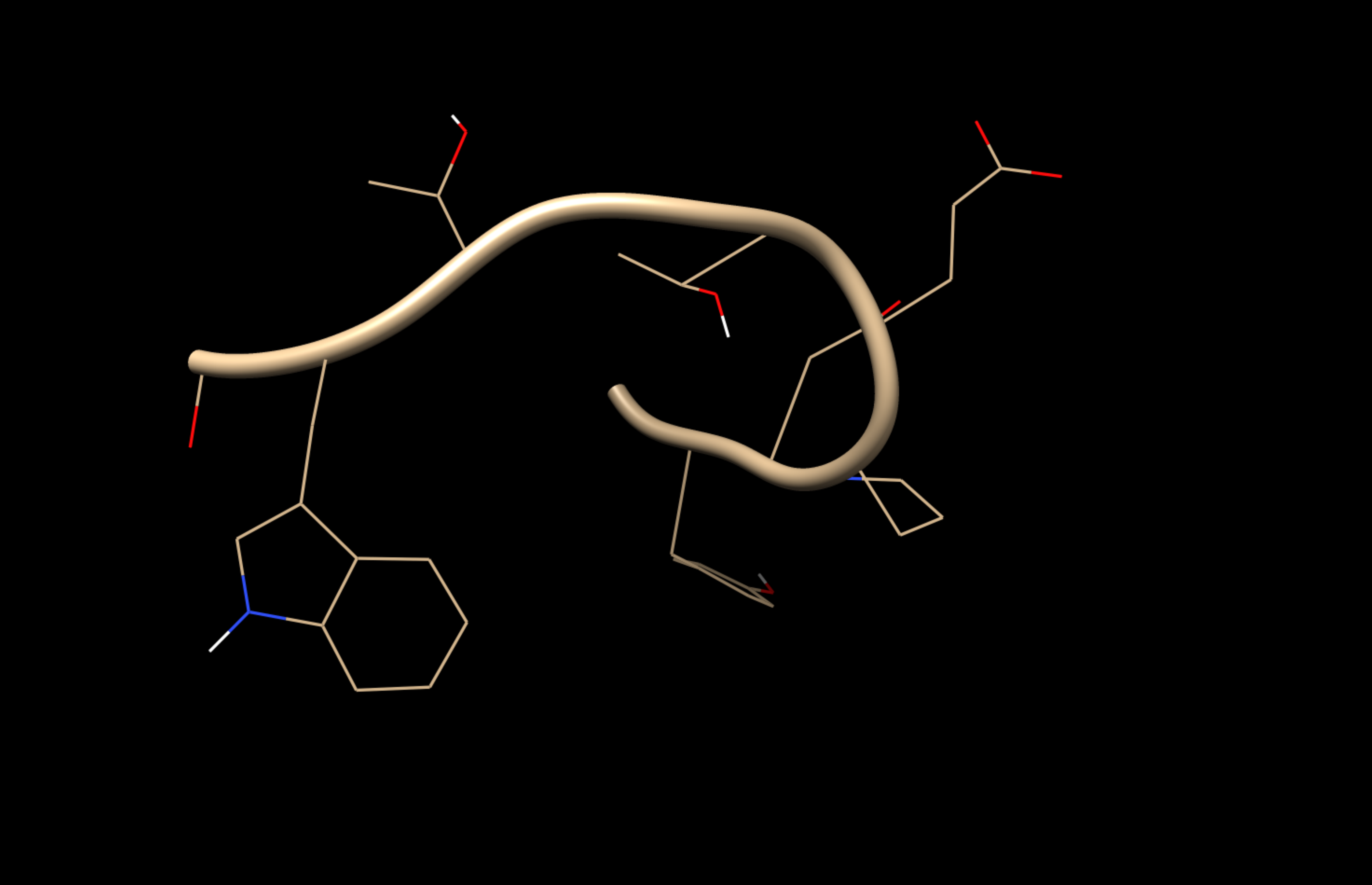}\label{sample_unfold}}
    \subfigure[distance map at t=160ns]{\includegraphics[width=.3\textwidth]{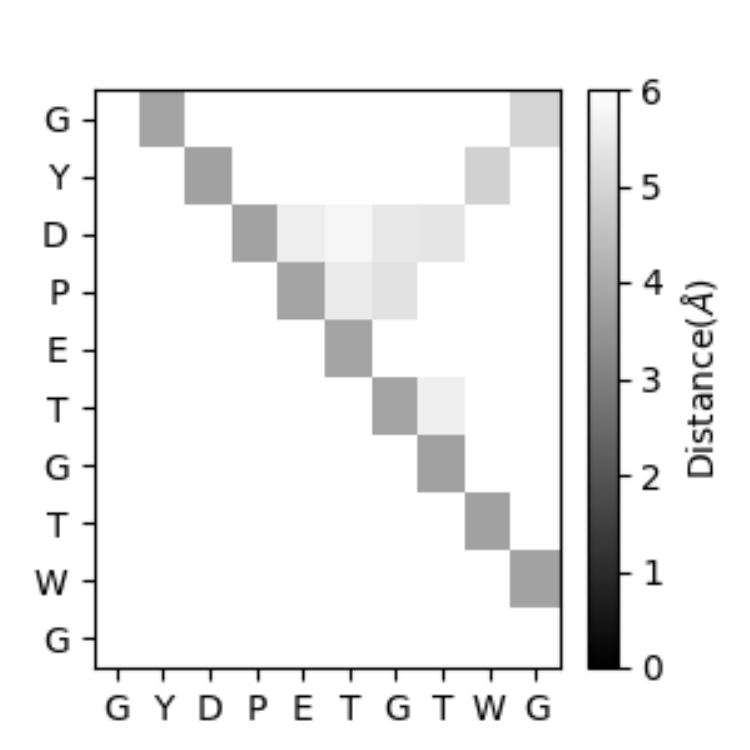}\label{dist_native}}
    \subfigure[distance map at t=120ns]{\includegraphics[width=.3\textwidth]{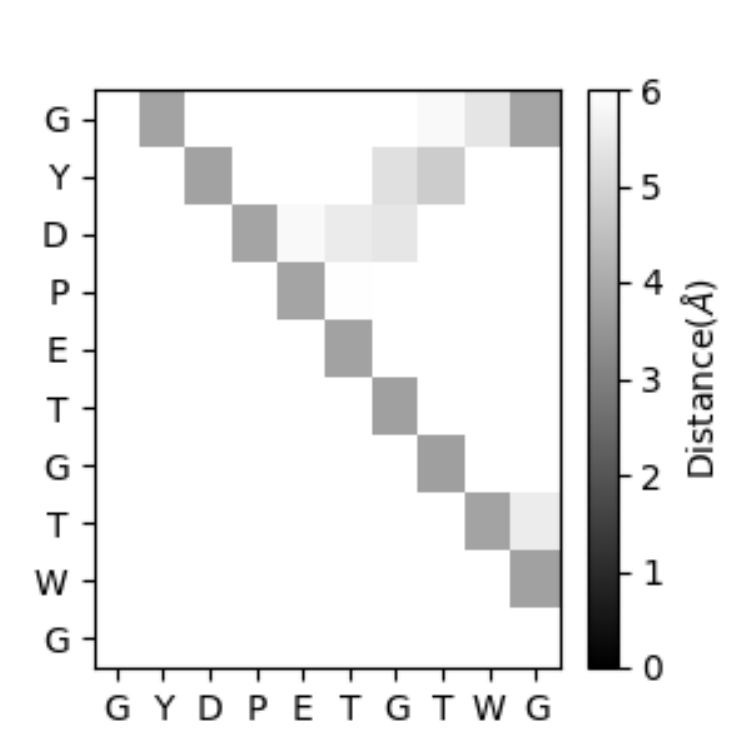}\label{dist_misfold}}
    \subfigure[distance at t=60 ns]{\includegraphics[width=.3\textwidth]{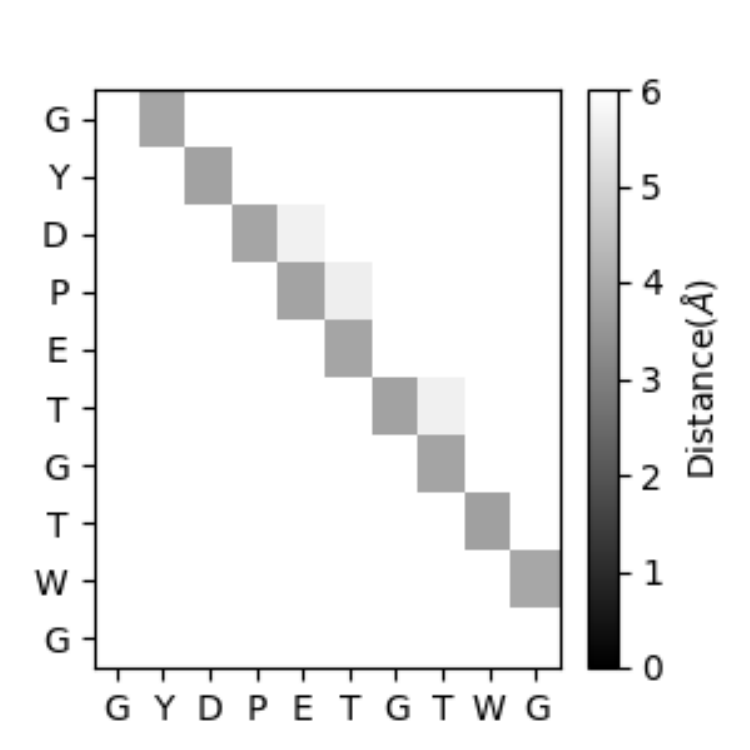}\label{dist_unfold}}
    \subfigure[barcode plot at t=160ns]{\includegraphics[width=.3\textwidth]{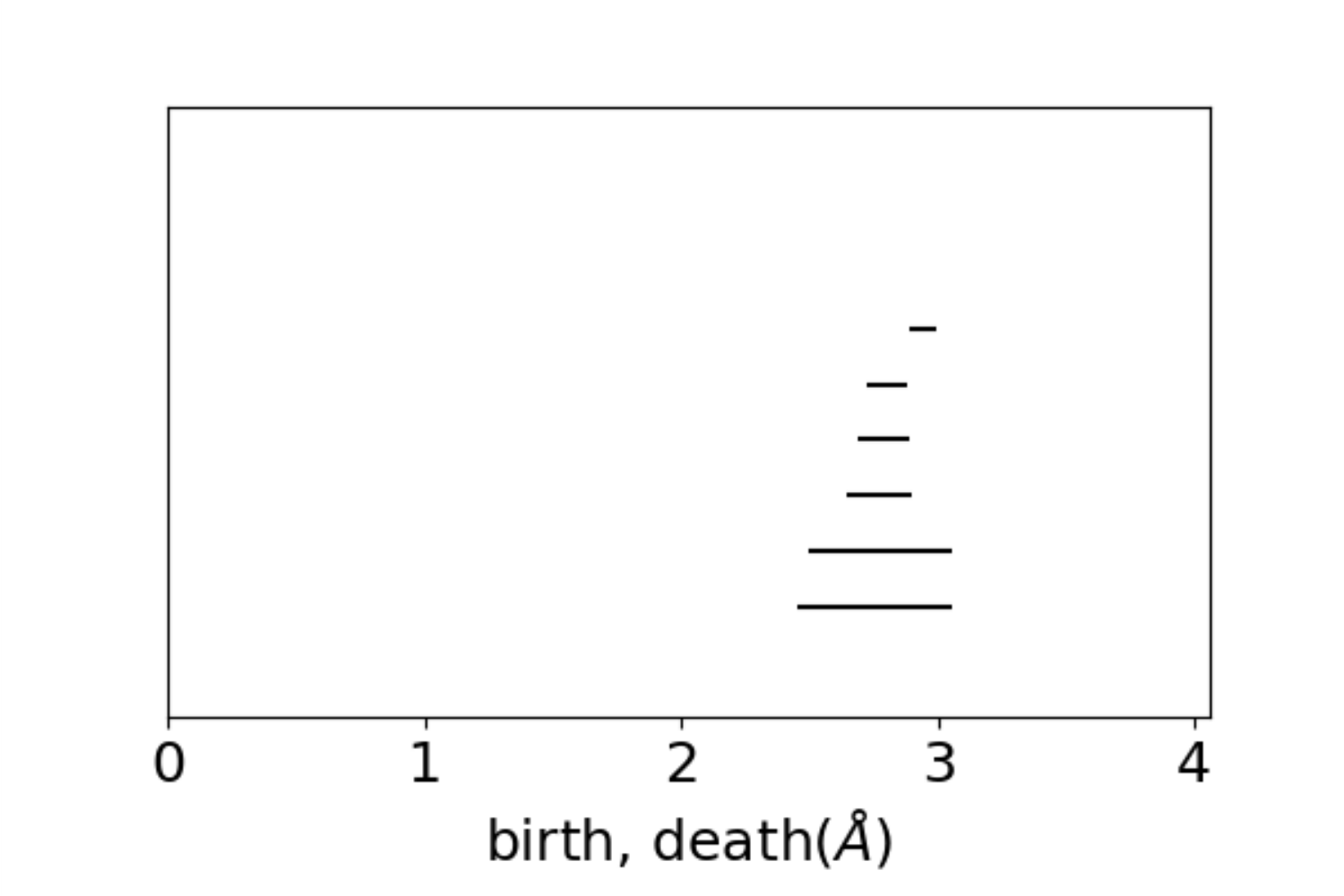}\label{bc_native}}
    \subfigure[barcode plot at t =120ns]{\includegraphics[width=.3\textwidth]{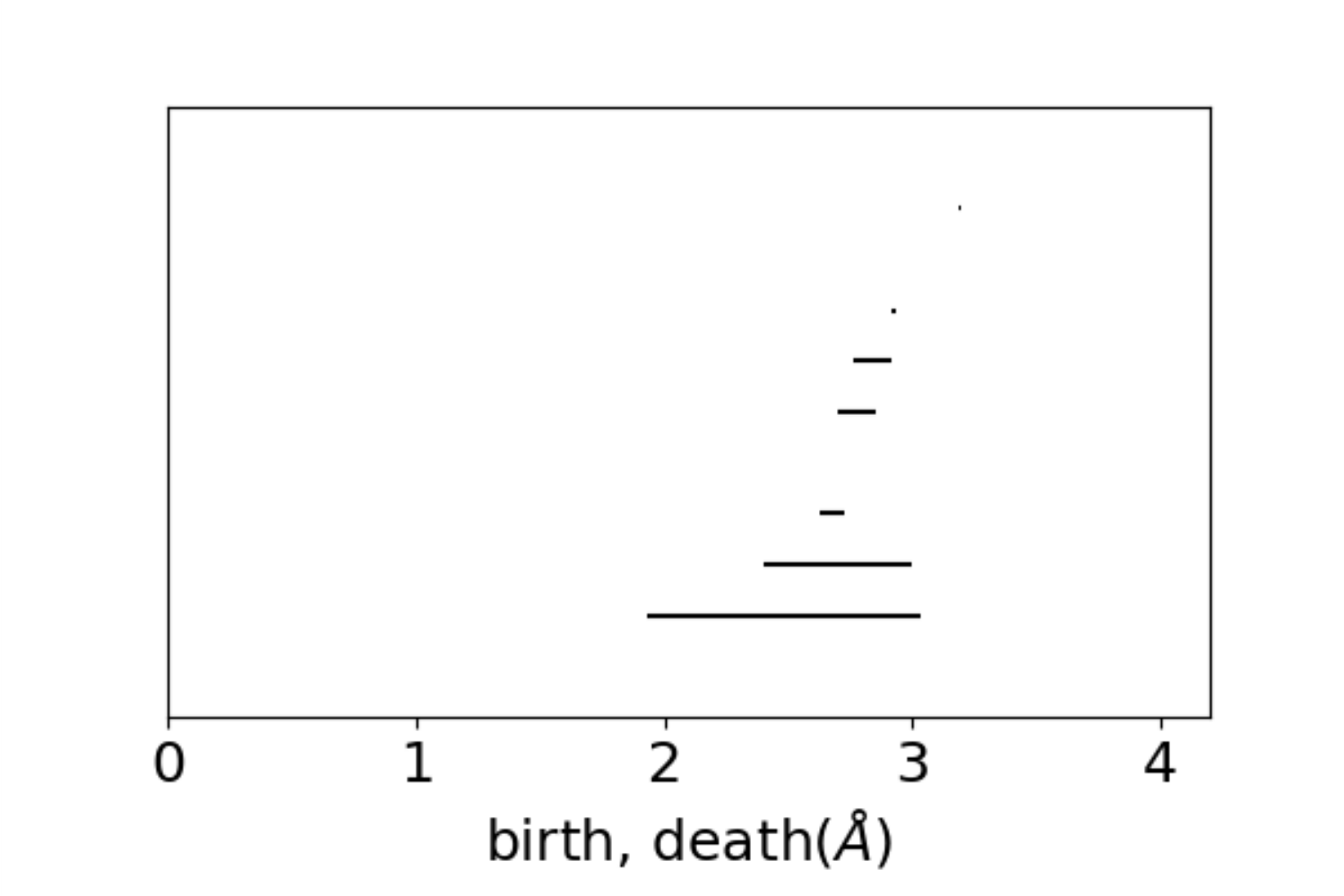}\label{bc_misfold}}
    \subfigure[barcode plot at t=60ns]{\includegraphics[width=.3\textwidth]{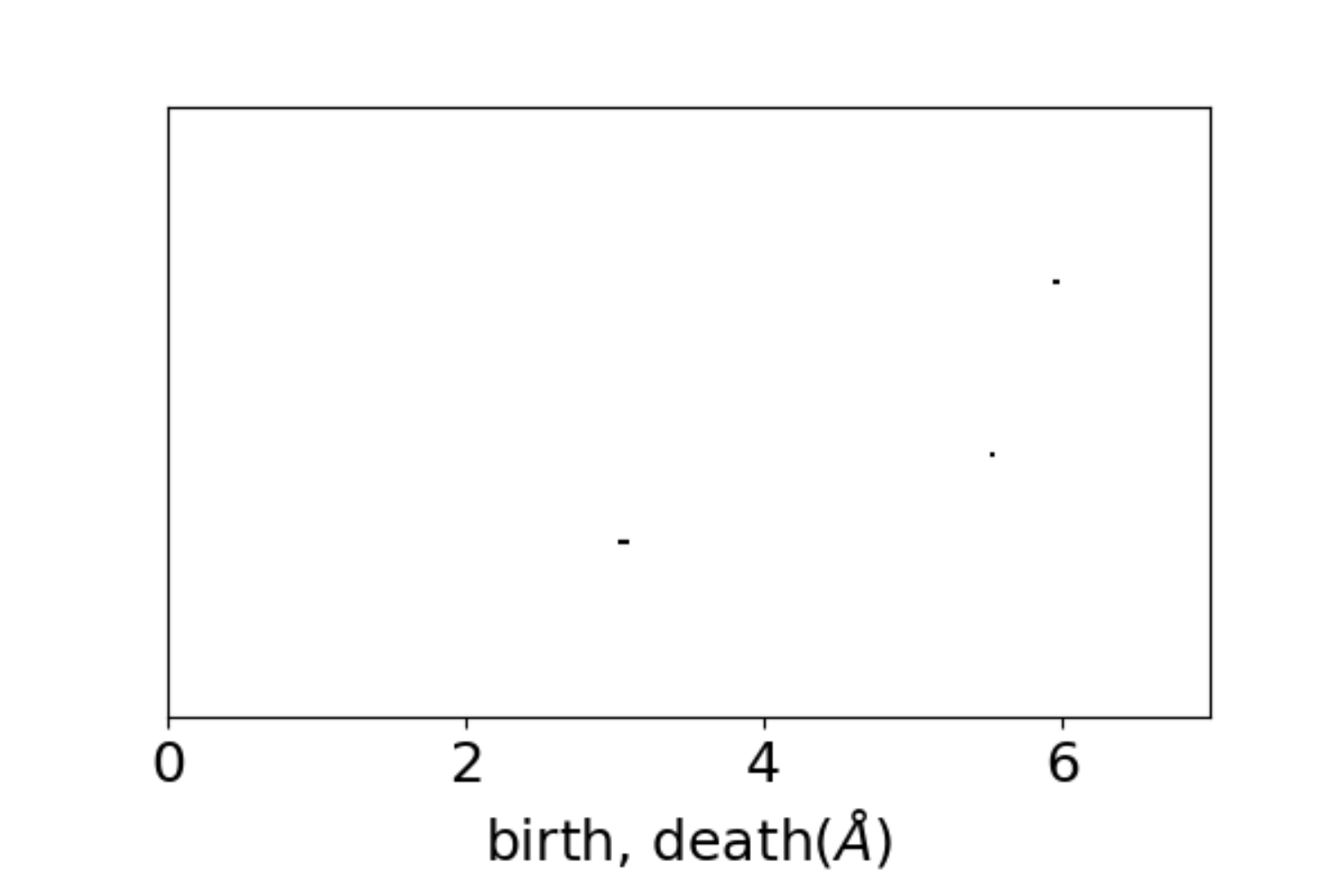}\label{bc_unfold}}
    \caption{(a)-(c): Example of protein configuration in native, misfolded and unfolded state. 
    (a) native state, $t=160$ns, $(h_1, h_2) = (22.7, 3.03)$. (b) misfolded state, $t=120$ ns, $(h_1, h_2) = (3.81,23.1)$. (c) unfolded state, $t=60$ns, $(h_1, h_2) = (3.60,6.44)$. (d)-(f): examples of distance maps for native (d), misfolded (e), and unfolded states (f). (g)-(i): persistent barcode plots for native (g), misfolded (h), and unfolded state (i). }
    \label{fig:example-configuration}
\end{figure}

This hypothesis was supported by the molecule snapshots.
In Fig.~\ref{fig:example-configuration}, we plotted examples of the chignolin configuration in the native, misfolded, and unfolded structures. This figure presents results consistent  with our hypotheses. Figs.\ref{dist_native}--\ref{dist_unfold} show the distance between amino acids in each state. In this plot, we calculated the distance between C$_\alpha$ atoms and inferred that this was the "distance" between corresponding amino acids. In the folded state, the distance between Tyr2--Trp9 is small, while the distance between Tyr2--Thr8 is small in misfolded states. We also noted that both the distances between Tyr2--Trp9 and Tyr2--Thr8 are small enough to assume these amino acids are "contacted," as discussed in the next subsection. In the unfolded state, the distance map shown in Fig.\ref{dist_unfold} has no clear structure.

\begin{figure}
    \centering
    \subfigure[$h_1$ and accumulated bar length]{\includegraphics[width=.45\textwidth]{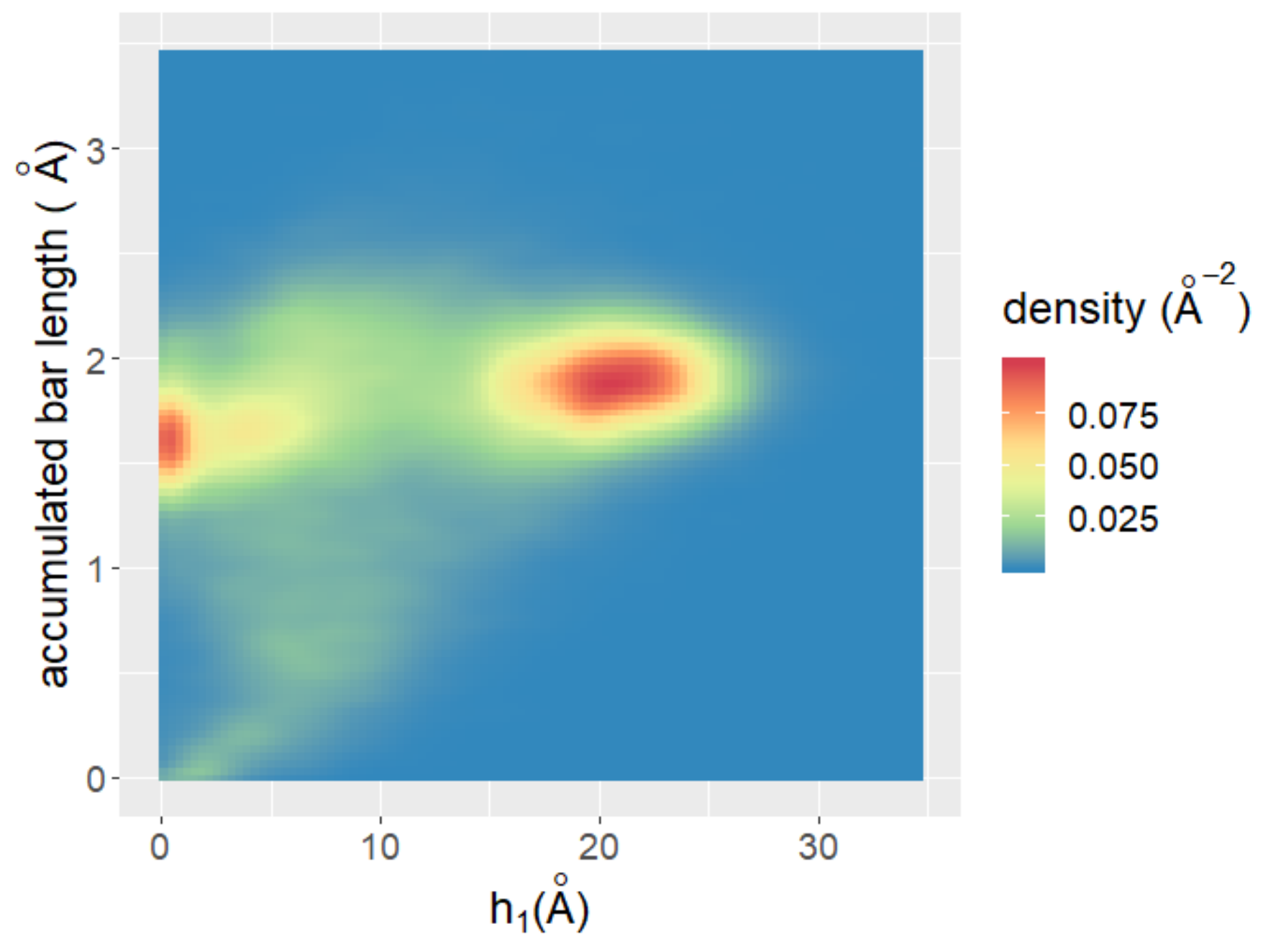}}
    \subfigure[$h_2$ and accumulated bar length]{\includegraphics[width=.45\textwidth]{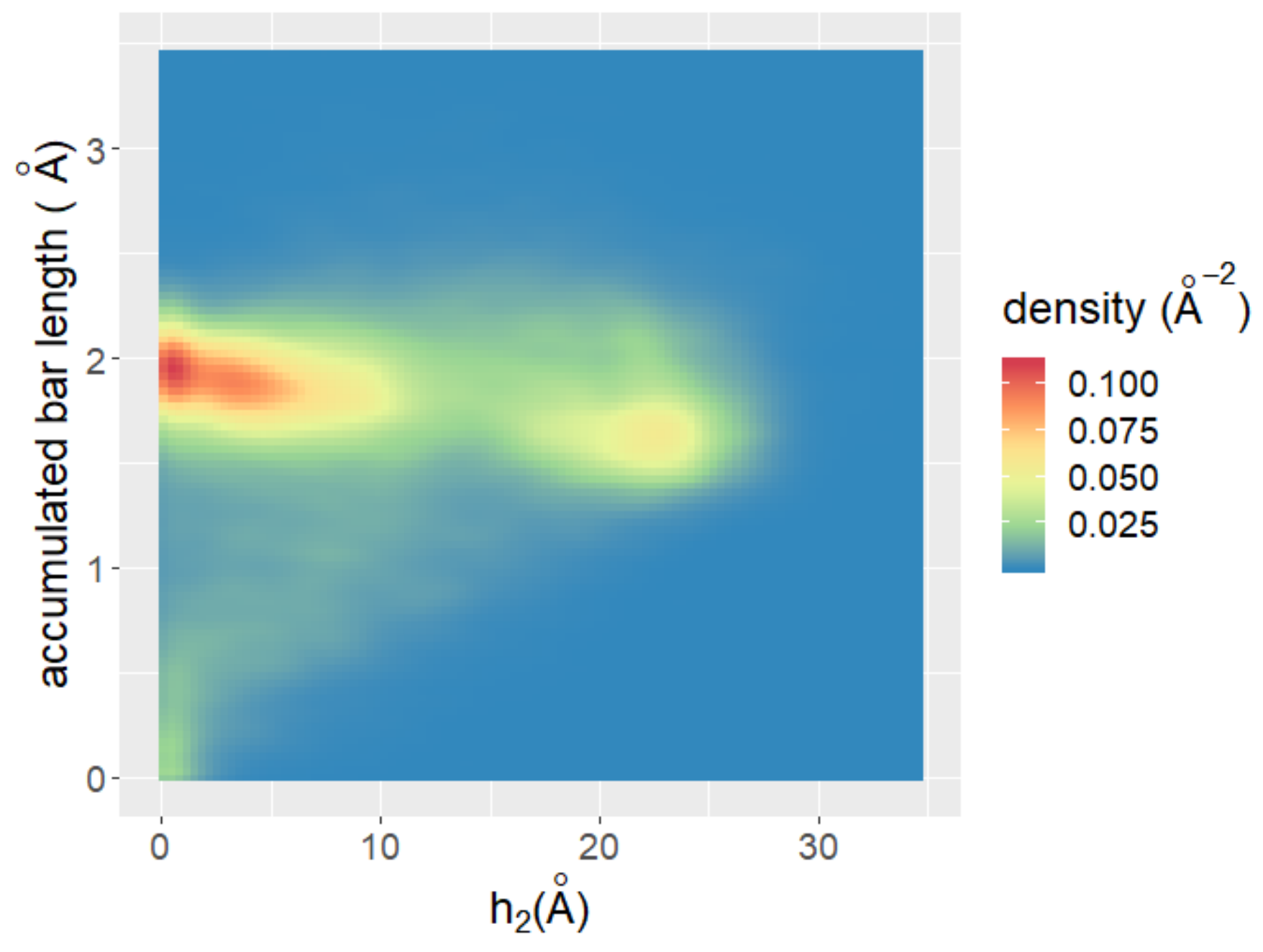}}
    \caption{Density plot of accumulated bar length and $h_i$'s.}
    \label{fig:accumulated_bar_length}
\end{figure}

\begin{figure}
    \centering
    \subfigure[cycles at t=160 ns]{\includegraphics[width=.4\textwidth]{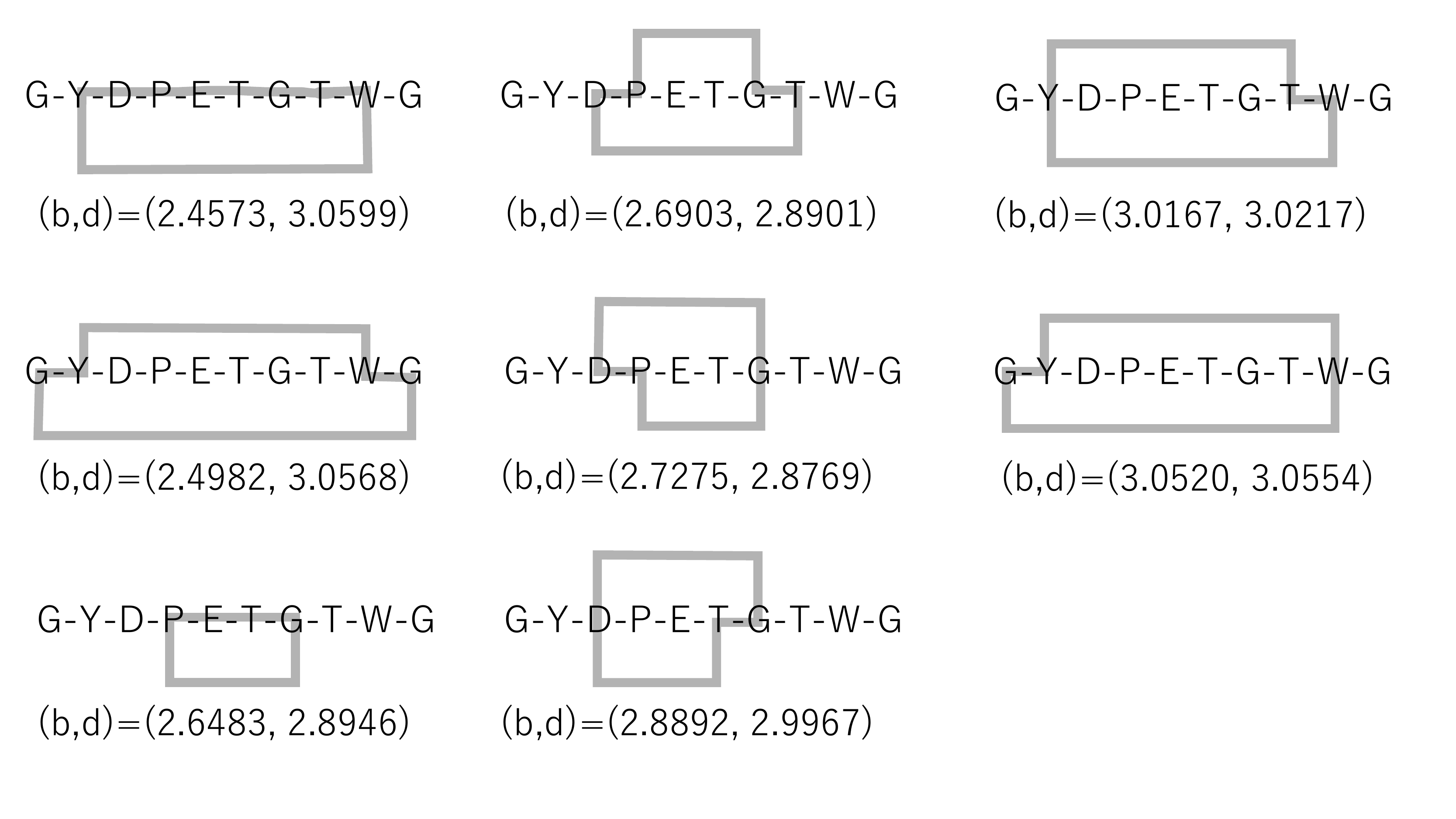}\label{cycle160}}
    \hfill
    \subfigure[cycles at t=120ns]{\includegraphics[width=.4\textwidth]{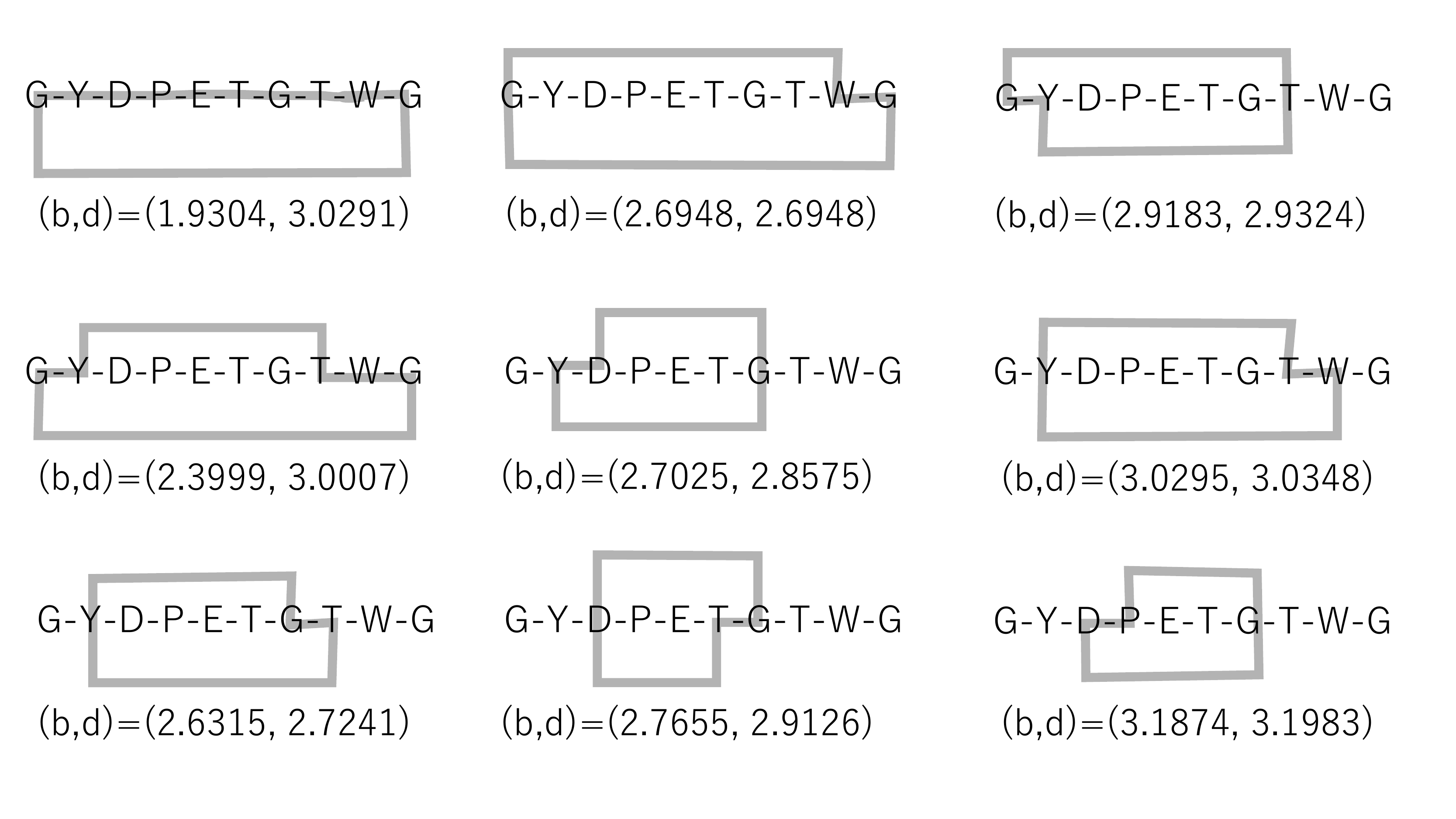}\label{cycle120}}
    \caption{Volume optimal cycles at native ($t=160$ ns) and misfolded ($=120$ ns) states. Gray lines indicate the edge included in the volume optimal cycle. (b,d) represent the birth and death time of each cycle, respectively. The unit of b and d are \AA.}
    \label{fig:cycles}
\end{figure}

Finally, we compared the TFV-based analysis and other PH analysis.
When applying PH to proteins, Xia and Wei proposed the molecular topological fingerprint (MTF) method\cite{Xia2014, Xia2015MultidimensionalData}, and claimed that "accumulated bar length" of persistent barcodes are useful in identifying  protein structure. To investigate the MTF, we plotted a "barcodes" diagram for the sample data in Fig.~\ref{bc_native}-\ref{bc_unfold}. In the barcode plot, each cycle is represented by a horizontal line, which begins at birth and ends at death. In this plot, cycles are sorted in ascending birth order. At $t=160$ ns, we have 8 loops, however, several loops had life times too short for observation in Fig.~\ref{bc_native}. Fig.\ref{bc_misfold} and \ref{bc_unfold} represent examples of barcodes for misfolded and unfolded states, respectively. In the misfolded state, we observed 9 cycles, however, it was difficult to distinguish the native and misfolded states. We observed 4 loops with very short lifetimes in the unfolded state, which can be easily distinguished from other states. Xia and Wei proposed to use accumulated bar length, i.e. the sum of bar length of all cycles, to describe the structure of protein. In Fig.~\ref{fig:accumulated_bar_length}, we show the density plot of $h_1, h_2$, and accumulated bar length. Clearly, the accumulated bar length for both clusters was approximately 1.5---2.0 \AA. Thus, we cannot distinguish these two peaks by accumulated bar length. 
The advantage of TFV analysis compared with other PH analyses relies on the volume optimal cycles, which offer much more information than persistent barcodes. For example, we show samples of  volume optimal cycles for native and misfolded states in Fig.\ref{fig:cycles}. From the list of native state cycles shown in Fig.\ref{cycle160}, we found that, as the amino acid radius of $r$ increases, the edge Tyr2-Tr9 emerge first, and by increasing $r$ further, the Gly1--Gly10, Pro4-Gly7, Asp3--Thr8, Asp3--Gly7, Asp3--Thr6, Tyr2--Thr8, and Gly1--Trp9 edges emerge, in this order. Compared with the MTF for the 2JOX protein presented by Xia and Wei\cite{Xia2014}, the chignolin's MTF is more complex. For 2JOX, the edges emerged between the closest adjacent amino acids, with one amino acid having only one edge in contact with the another strand of the $\beta$-sheet. In our case, several  amino acids contact two or more neighbors. For example, Gly1 contacts Trp9 and Gly10, while Asp3 contacts  Gly7, Thr6, and Thr8. The list of volume optimal cycles in Fig.~\ref{cycle160} also showed that 4 of 8 cycles include the edge Tyr2--Trp9. However, in the misfolded state at $t=120$ ns, we observed 4 of 9 cycles included the edge Tyr2--Thr8, as shown in Fig.~\ref{cycle120}. These observations are consistent with the result presented in Fig.~\ref{fig:nmf_components}. We also noted that births and deaths are not sufficient to distinguish the cycles in the native and misfolded states. For example, cycles Gly1--Tyr2--Trp9--Gly10 in Fig.~\ref{cycle160} and Gly1--Tyr2--Thr8--Gly10 in Fig.~\ref{cycle120} have nearly the same births and deaths. The volume optimal cycle reveals the difference between these "similar" generators.

\subsection{Comparison with the cartesian coordinates and contact-map results} 

It was instructive to investigate the relationship between our TFV analysis and other previous methods. Here, we compare our result with those of cartesian coordinate and contact map-based analysis.

\begin{figure}
    \centering
    \subfigure[]{\includegraphics[width=.3\textwidth]{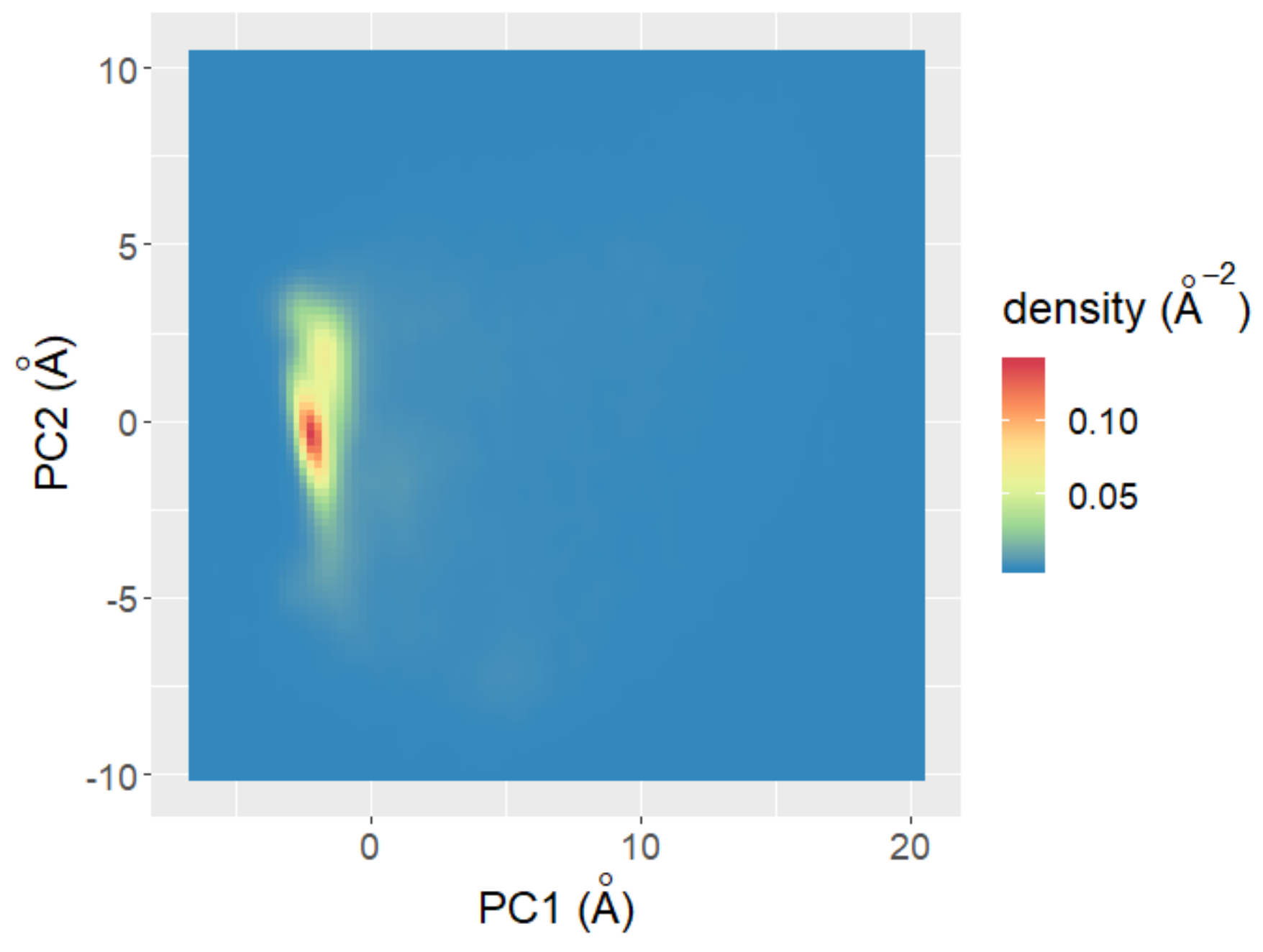}\label{pca_12}}
    \subfigure[]{\includegraphics[width=.3\textwidth]{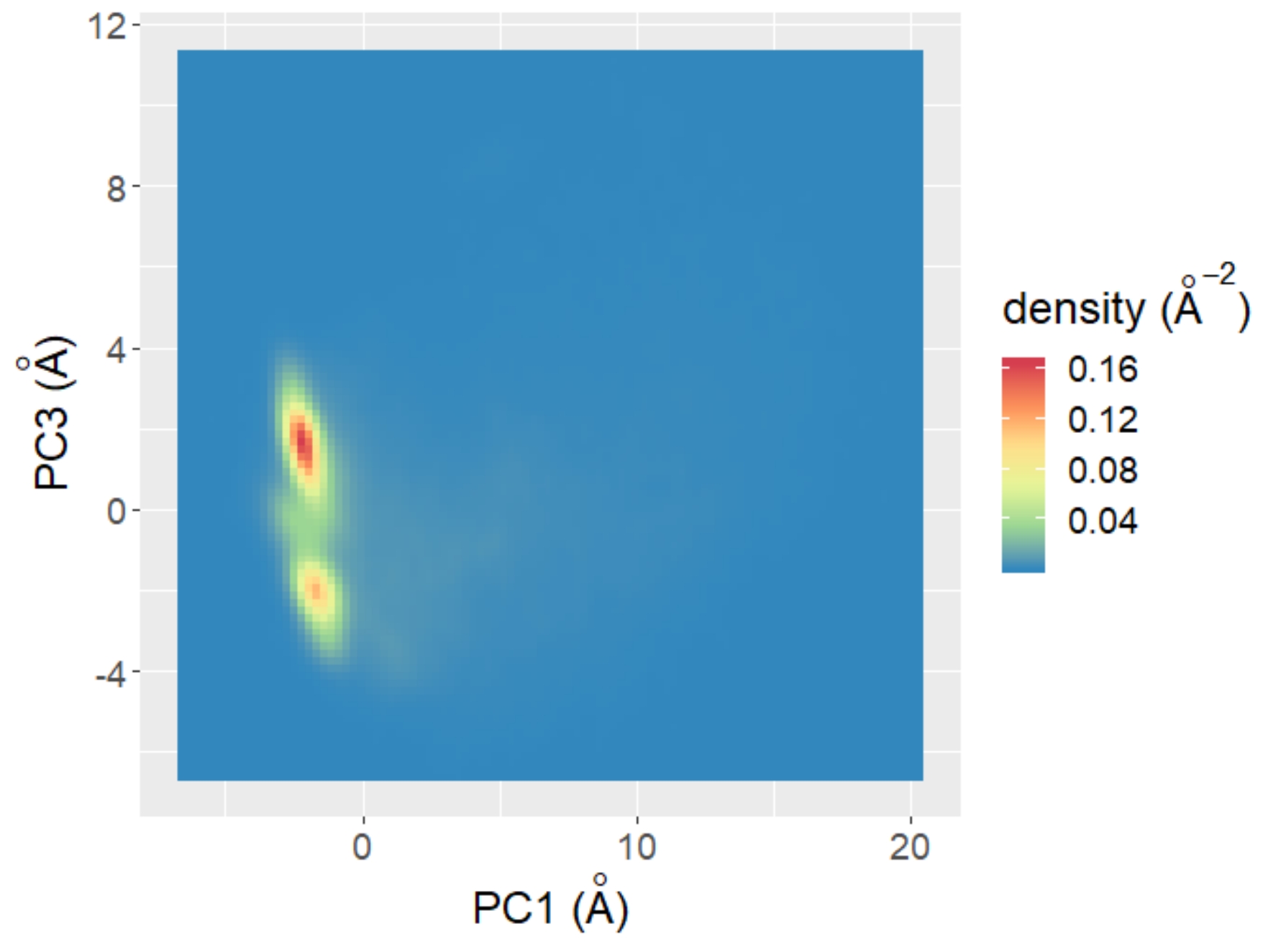}\label{pca_13}}
    \subfigure[]{\includegraphics[width=.3\textwidth]{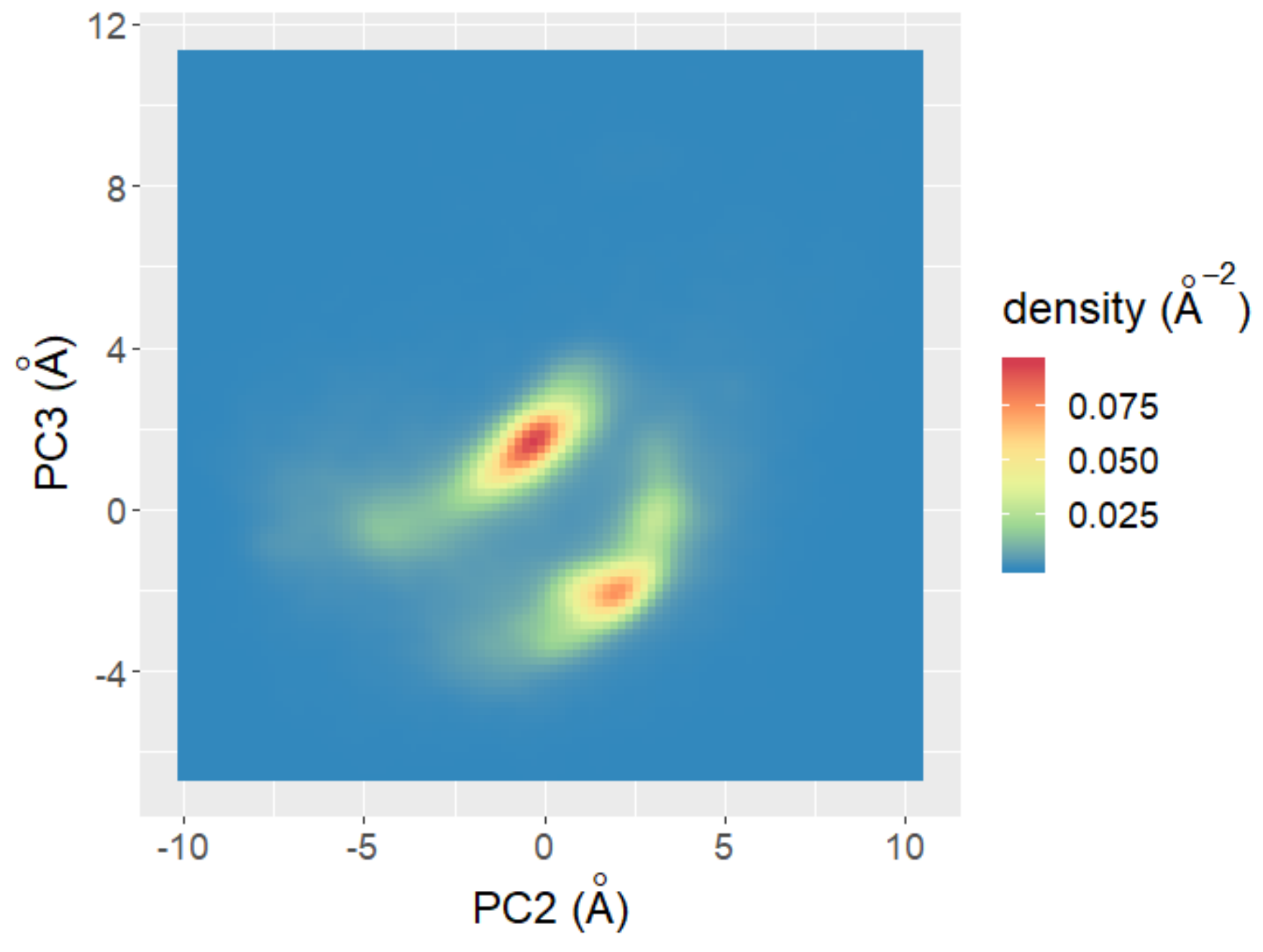}\label{pca_23}}
    \caption{The density plots in the reduced space obtained by principal component analysis based on cartesian coordinates of C$_\alpha$ atoms. (a) 1st and 2nd components. (b) 1st and 3rd components. (c) 2nd and 3rd components.}
    \label{fig:traditional_pca}
\end{figure}

First, we compared our results and those based on the cartesian coordinates of C$_\alpha$ atoms, as described by Mitsutake and Takano\cite{doi:10.1063/1.4931813}. In this analysis, we constructed the $3n$-dimensional vector $\bm{R} = (x_1, y_1, z_1, x_2, y_2, z_2, \cdots, x_n, y_n, z_n)$, where $(x_i,y_i,z_i)$ represents the position of $i$-th C$_\alpha$ atom. We then reduced dimensionality by PCA. NMF is not available in this case as cartesian coordinates can be negative. The density plots in the reduced space are represented in Fig.~\ref{fig:traditional_pca}. We identified two clear peaks of density in Fig.~\ref{pca_23}. These results are consistent with the result of Mitsutake and Takano\cite{doi:10.1063/1.4931813}. Importantly, the first principal component does not contribute to the cluster identification. These results indicate that the PCA is strongly affected by structural changes which are not related to the transition between the folded and misfolded states. A possible explanation is the large degree of freedom in the unfolded state. In unfolded states, most amino acids can move freely, which results in large cartegian coordinate fluctuations. In TFV analysis, we had few cycles in the unfolded state, and the corresponding TFV became nearly $0$. Therefore, we were able to avoid this large fluctuation in the unfolded state by using TFV.

\begin{figure}
    \centering
    \subfigure[color=$h_1$]{\includegraphics[width=.45\textwidth]{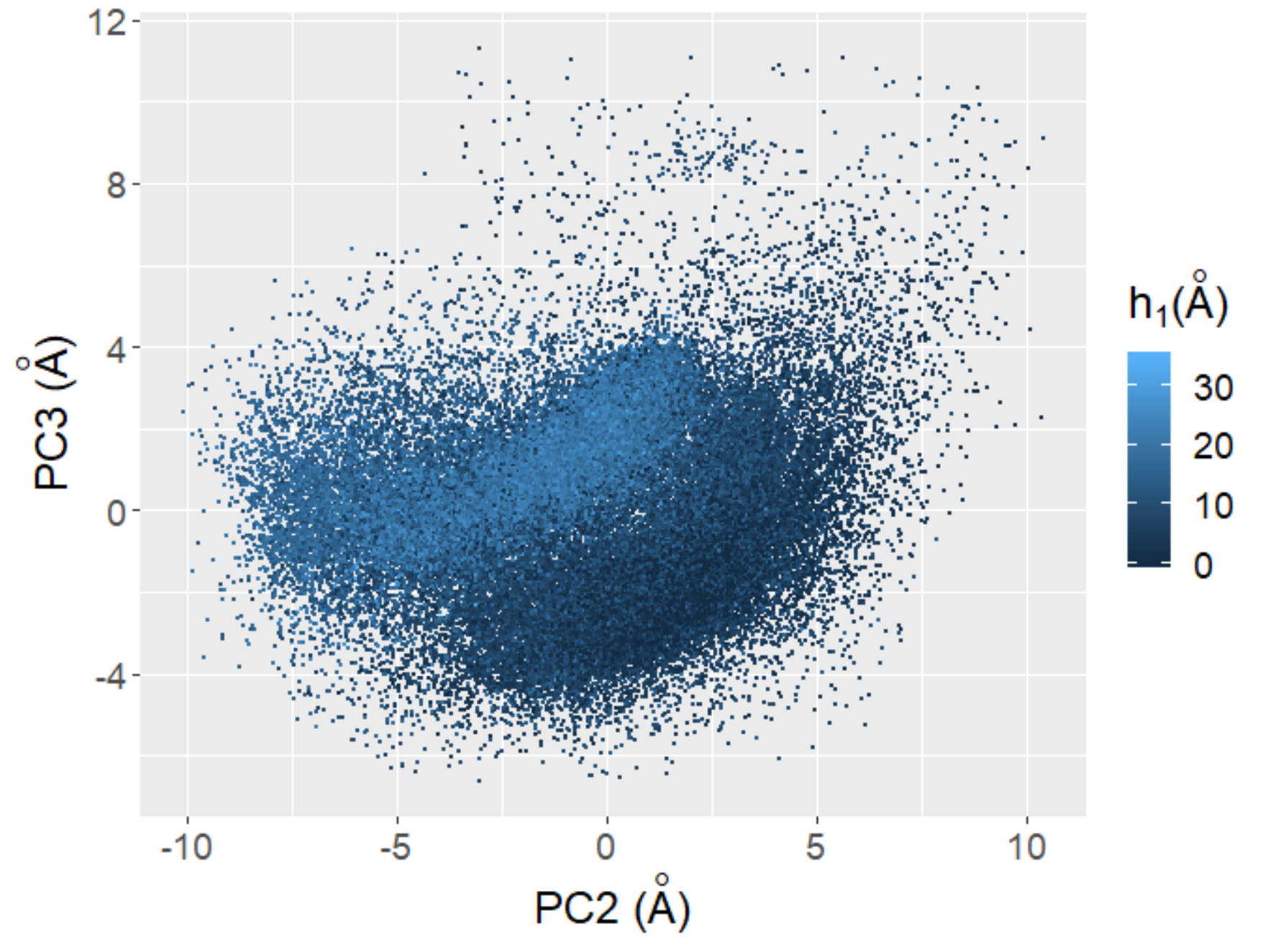}}
    \subfigure[color=$h_2$]{\includegraphics[width=.45\textwidth]{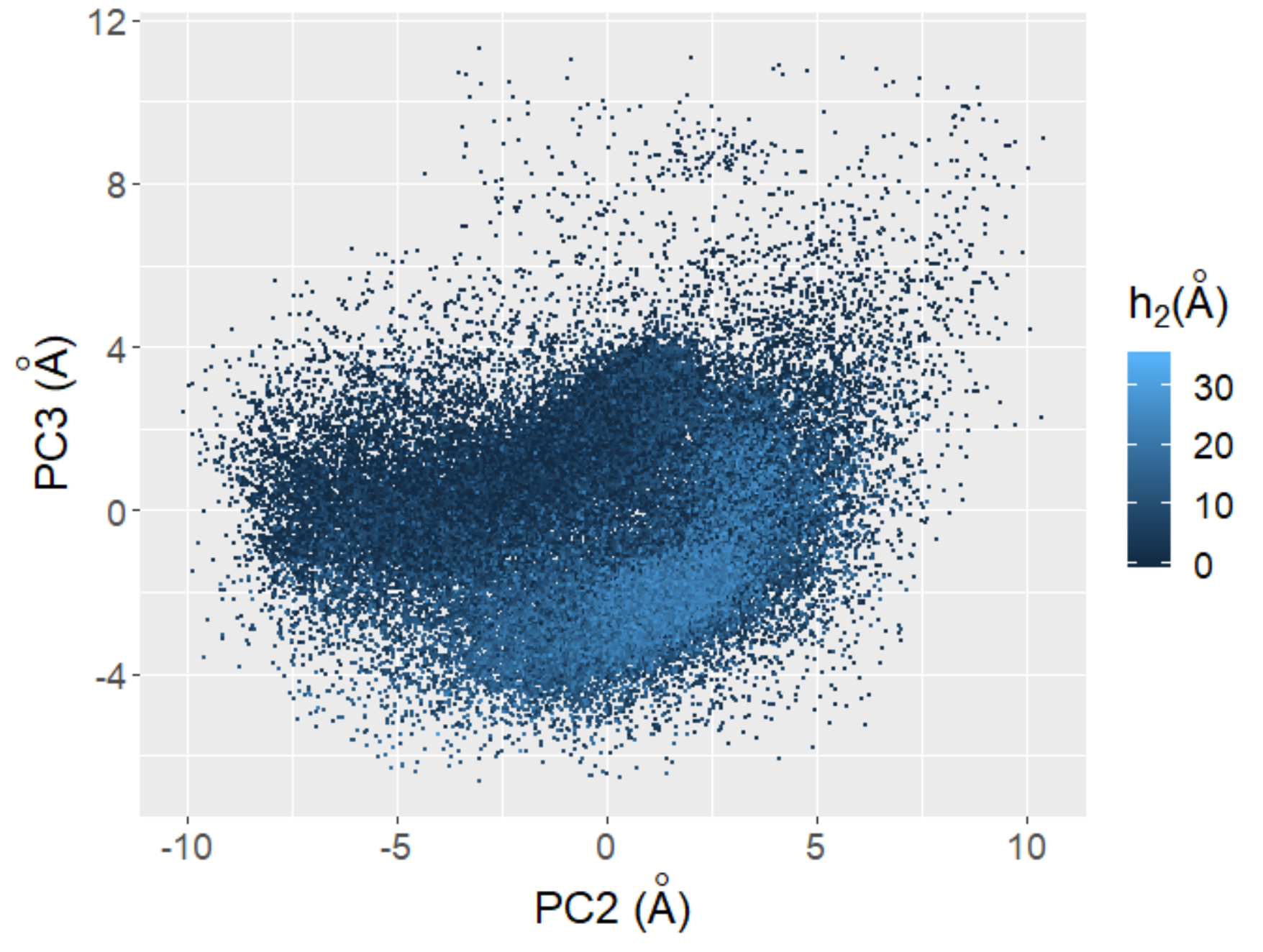}}
    \caption{Scatter plot in second and third cartesian coordinate principal components. Color indicates the values of (a)$h_1$ and (b) $h_2$ obtained by TFV NMF, respectively.}
    \label{fig:PCA_nmf}
\end{figure}

To confirm the cluster consistency between the TFV-based analysis and cartesian coordinate-based analysis, we investigated the relationship between the $h_i$ in Fig.~\ref{density_nmf} and the PCA result. Fig.~\ref{fig:PCA_nmf} shows the scatter plot of the second and third principal components, where the point color indicates the score $h_1$ and $h_2$ obtained by NMF. Comparing this figure with Fig.~\ref{pca_23}, $h_1$ returns large values for the cluster PC3 $\sim 0$, while $h_2$ are large for the cluster at PC3 $\sim -4$. Therefore, we concluded that the clusters obtained by cartesian coordinate PCA and TFV-based analysis coincide.

Next, we conducted PCA and NMF analysis based on contact maps using the method proposed by Ernst {\it et al.}\cite{Ernst2015Contact-Dynamics}. In this method, we calculated the distance between $C_{\alpha}$ atoms and created the contact map $D_{ij}$ as 
\begin{equation}
    D_{ij} =
    \begin{cases}
    1 & \mbox{if the distance between $i$-th and $j$-th C$_\alpha$ is smaller than 8 \AA }\\
    0 & \mbox{otherwise}
    \end{cases}
    \label{eq:Dij}
\end{equation}

\begin{figure}
    \centering
    \subfigure[result of PCA ]{\includegraphics[width = .45\textwidth]{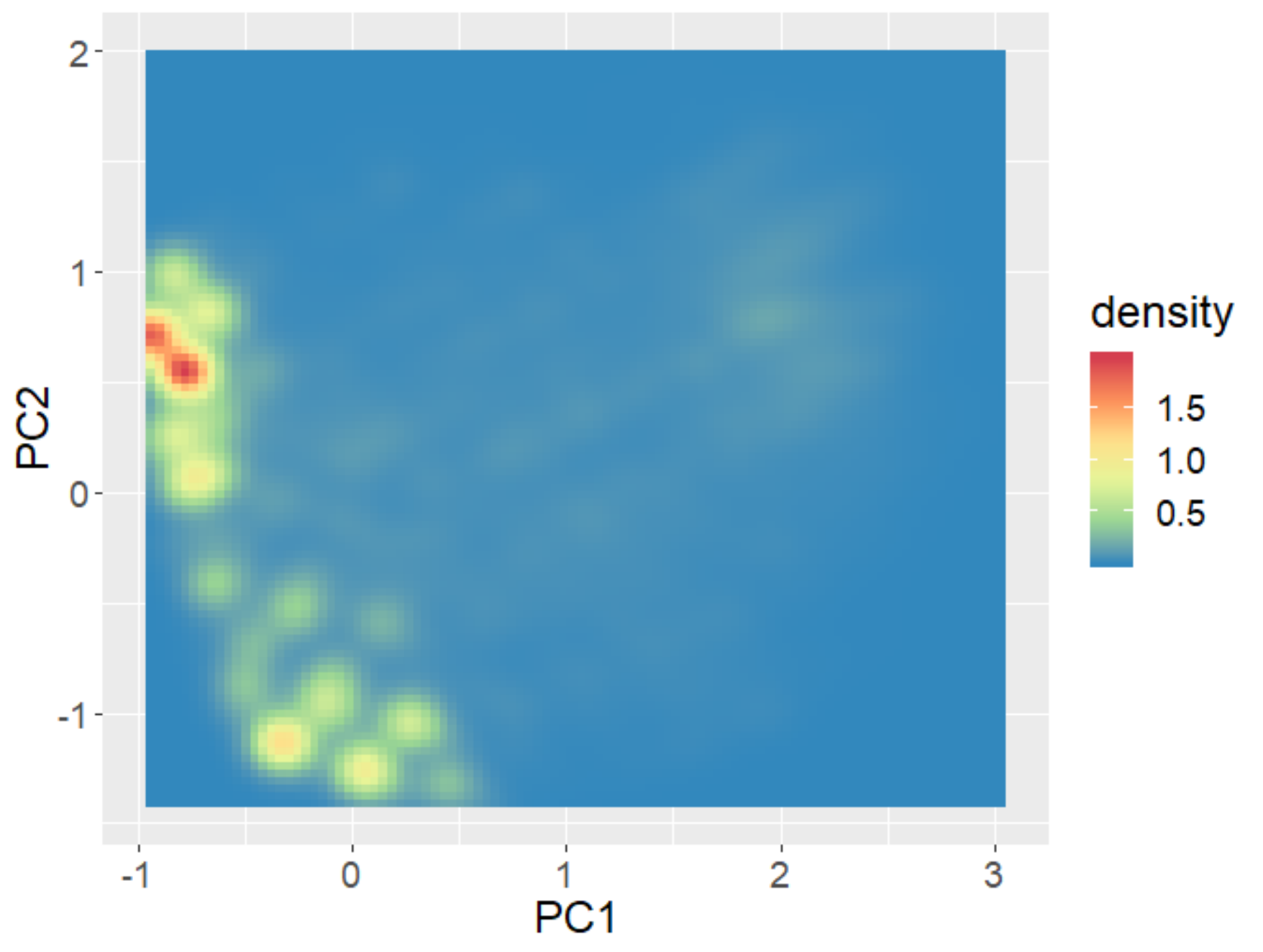}\label{contact-PCA}}
    \subfigure[result of NMF ]{\includegraphics[width =.45\textwidth]{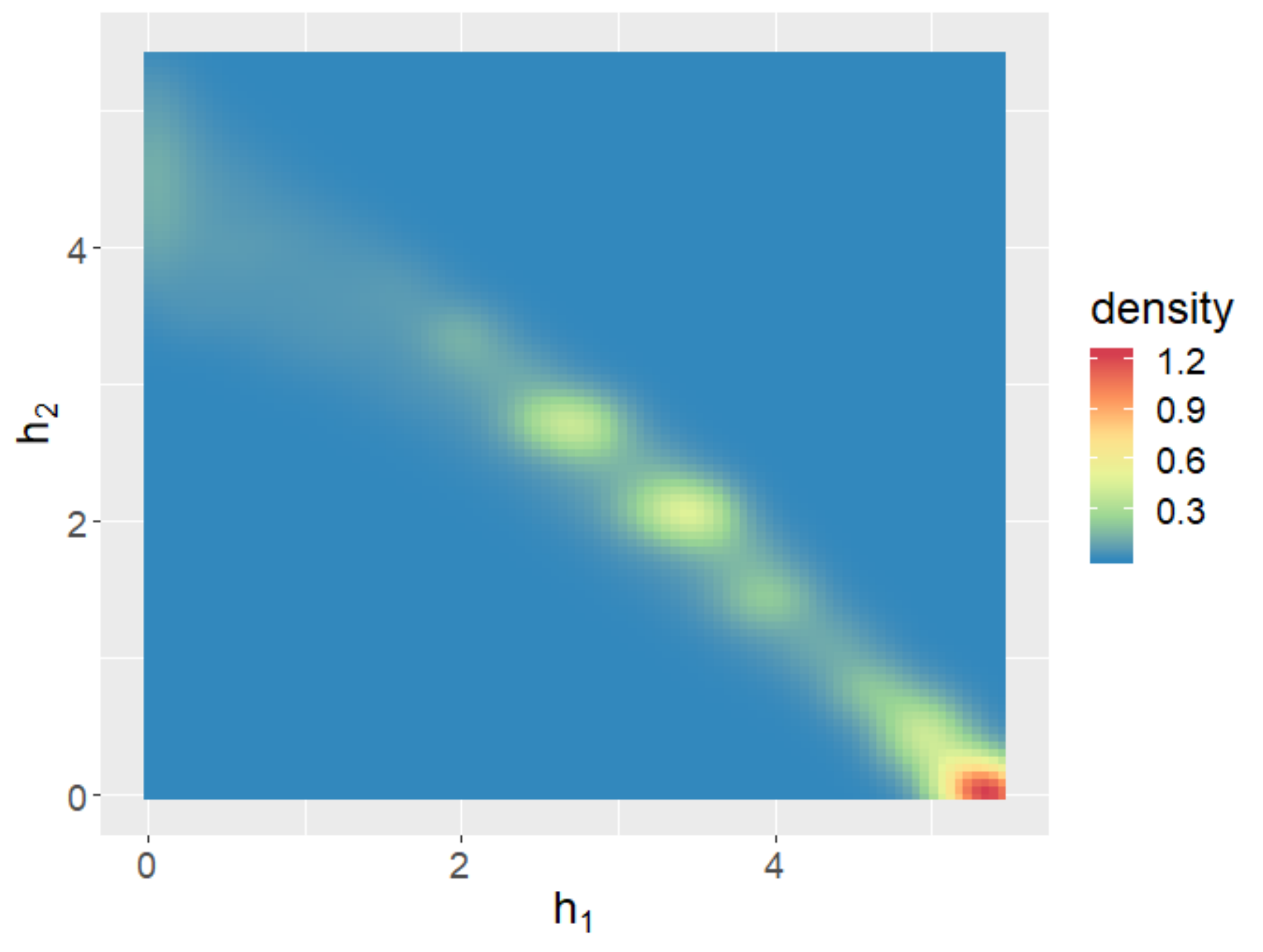}\label{contact-NMF}}
    \caption{Result of PCA and NMF of contact map $D_{ij}$}
    \label{fig:contact}
\end{figure}

$D_{ij}$ is non-negative, so we could apply both PCA and NMF. 
In Fig.\ref{fig:contact}, we show the density plot obtained by PCA and NMF of $D_{ij}$. In both cases, it was difficult to identify the folded or misfolded states.
Compared with the analysis based on contact maps, TFV selects the "shortest" distance automatically. When the distance between Tyr2--Tyr8 and Tyr2--Trp9 are both shorter than 8 \AA, $D_{ij}=1$ for both pairs. In this case, we cannot distinguish the folded and misfolded states by contact map. Of 100,000 samples, 80,366 samples showed that both Tyr2--Tyr8 and Tyr2--Trp9 are "contacted" and we failed to distinguish the native and misfolded states. However, TFV depends on the distance difference; if the distance between Tyr2--Trp9 is smaller than that between Tyr2--Tyr8, the edge between Tyr2--Tyr8 is not included in the volume optimal cycle since at the birth of the cycle including Tyr2--Tyr8, Tyr2--Trp9 is not "contacted." In other words, TFV is sensitive to the difference in distance between atoms, while contact maps are sensitive to the absolute distance. This explains why TFV successfully detected two clusters that were not captured by the contact map-based method.

Before concluding this subsection, we discuss a previous study by Mitsutake and Takano describing the structure of native and misfolded state\cite{doi:10.1063/1.4931813} in which they investigated the distance between many atom pairs and found that the hydrogen bond between Asp3 and Gly7 is strongly related to the difference between the native and misfolded states. Fig.~\ref{fig:nmf_components} shows that the contribution of the edge between Asp3 and Gly7 is larger in $\bm{w}_2$ than in $\bm{w}_1$, which is consistent with their results. However, our analysis showed that the edge between Tyr2 and Thr8 is more remarkable in $\bm{w}_2$, which is not mentioned previously. Thus, further studies are needed to determine the difference between our work and previous studies. 

\subsection{Dynamics in reduced space}

\begin{figure}
    \centering
    \subfigure[]{\includegraphics[width = .45\textwidth]{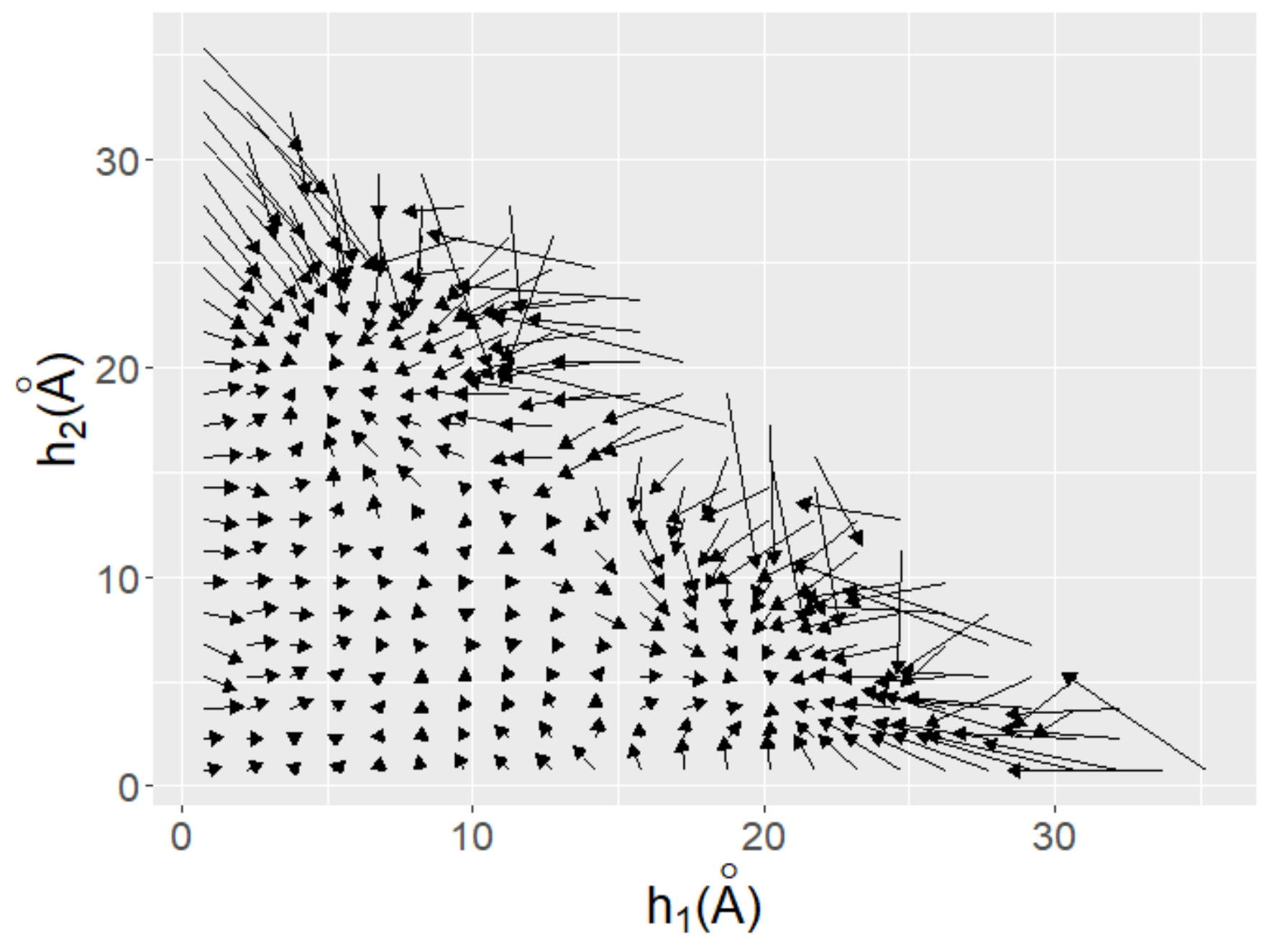}\label{fig:flow-global}}
    \subfigure[]{\includegraphics[width=.45\textwidth]{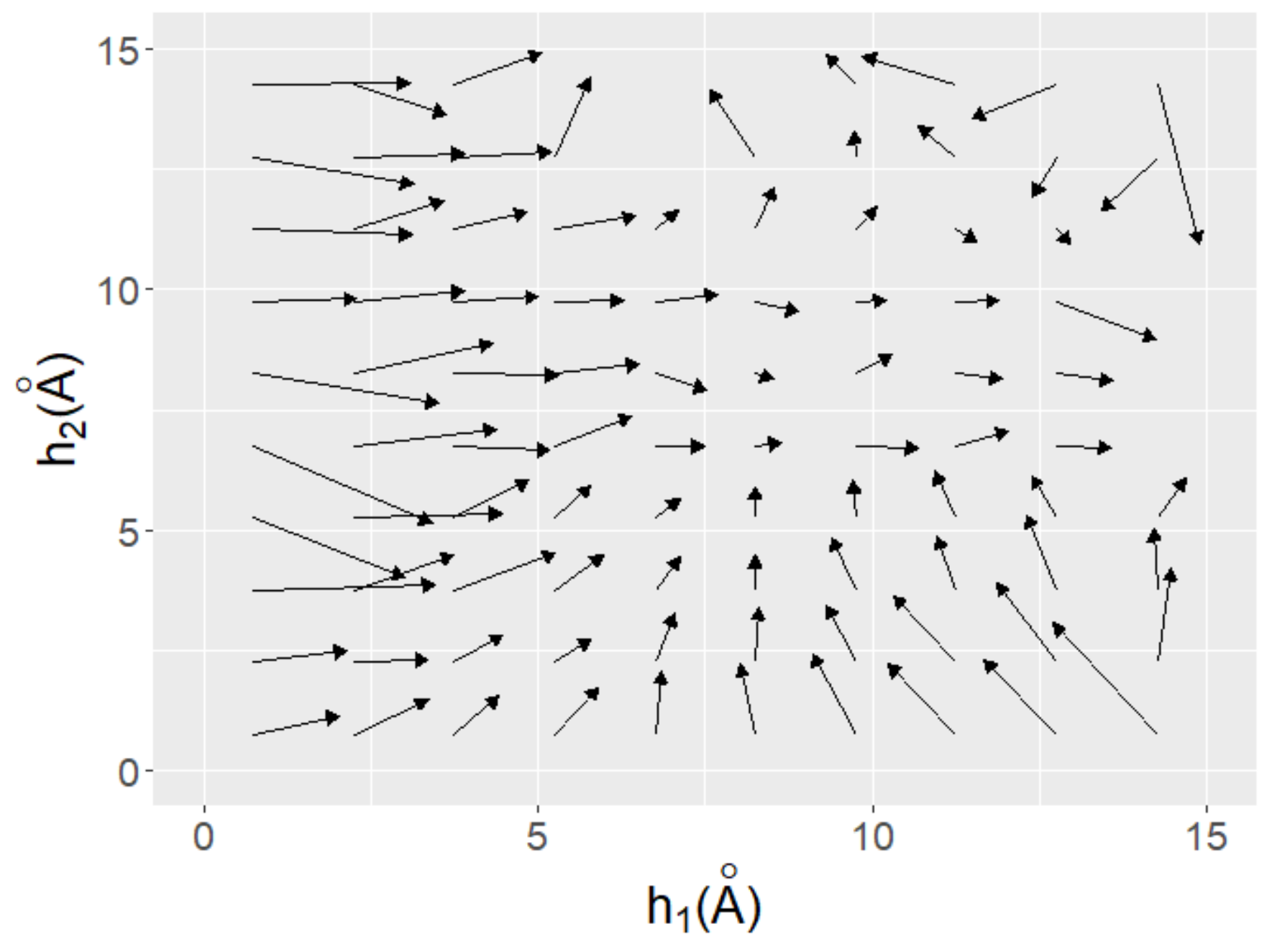}\label{fig:flow-local}}
    \caption{Average flow $\delta \bm{h}(t) = \bm{h}(t+\delta t)-\bm{x}(t)$. Left: flow in the entire two-dimensional space. Right: flows at $0 \text{\AA} \le h_1, h_2 \le 15 \text{\AA}$. \label{fig:flow}}
\end{figure}

To investigate the transition between native, misfolded and unfolded states,  we plotted the average flow in Fig.\ref{fig:flow}. First, we calculated $\delta\bm{h}(t)= \bm{h}(t+\delta t) -\bm{h}(t)$, where
$\delta t $ = 10ps. Next, we divided the two-dimensional space into grids of size $1.5$ \AA $\times 1.5$ \AA and calculated the average of $\delta \bm{h}(t)$ for each grid. Fig.~\ref{fig:flow-global} shows the flow in the entire two-dimensional space. The results indicate that there are two stable solutions at $(h_1, h_2) \sim (20 \text{\AA} ,5 \text{\AA}) $ and $(5 \text{\AA},20 \text{\AA})$, respectively. 
These two stable points correspond to folded and misfolded states, as discussed above. The positions of these fixed points differ slightly from the density peak, which is due to the constraint of $h_1, h_2 \ge 0$. Though the density peak is on the $h_1$ and $h_2$ axis, these points cannot be fixed point as any configuration change drives the system away from the axis.
We also found that the flow along the line $h_1 + h_2 \sim 25$ is strong, whereas the flow at small $h_1 + h_2$ is very weak. To investigate the dynamics in this area more clearly, we plotted the flow at $0 \text{\AA} \le h_1, h_2 \le 15 \text{\AA}$ in Fig.~\ref{fig:flow-local}. These results strongly suggest that there is a saddle point at $(h_1, h_2)\sim (12 \text{\AA},12 \text{\AA})$. Therefore, this position is likely the transition state. These results also demonstrate that there is no fixed point corresponding to the unfolded state. Once the chignolin molecule reaches the thermal noise-induced unfolded state, it remains unfolded for an extended period of time. However, this is not due to the attraction to the stable fixed point, but rather the slow dynamics in reduced space.

\section{Conclusion}\label{Sec:Conclusion}

In this study, we analyzed the chignolin folding process using PH and proposed TFV as a feature to characterize protein structure. TFV allows us to combine PH with machine learning methods, such as PCA or NMF. In particular, we show that NMF analysis of TFV provides essential information on folding dynamics. By investigating flow in the reduced space, we found that there are two stable fixed points that correspond to the native and misfolded states, one saddle point corresponding to transient state, and one unfolded state. The difference between the native and misfolded  states lies in the edge differences between Tyr2--Trp9 and Tyr2--Thr8. The unfolded state has no fixed point, while the protein remains unfolded for a long time due to the slow dynamics.

Our results show that featurization of protein structure by TDA is promising. Several previous studies attempted to apply TDA to biomolecule dynamics. For example, Yao {\it et al}. applied Mapper, a major TDA method, to investigate  RNA folding\cite{doi:10.1063/1.3103496}. Xia and Wei proposed MTF to characterize protein structure\cite{Xia2014, Xia2015MultidimensionalData}. Gameiro {\it et al.} investigated the relationship between PH and protein compressibility\cite{Gameiro2015}. Compared to these previous works, there are several advantages to the TFV-based analysis. First, TFV uses volume optimal cycles, which include significantly more information than does persistent barcodes. As shown in Sec.~\ref{Sec:Result}, volume optimal cycles can distinguish different structures that have same births and deaths. Another advantage is applicability to machine learning. As we discussed in the Introduction, the fluctuation in number of cycles causes difficulty when applying machine learning with PH. However, the TFV dimension only depends on the number of amino acids; therefore we can easily apply machine learning methods, such as PCA, NMF, and deep learning. Notably, TFV calculations require large computational cost. Therefore, it would be difficult to calculate TFV using all atoms in macromolecules.

The method developed in this study is powerful, however, further development is also possible. First, several approaches can be taken to construct TFVs from the volume optimal cycles. Here, we used the sum of the deaths as the edge weight, but we could alternatively use the sum of births, the product of deaths, or a combination of births and deaths. For chignolin, there is no qualitative difference when we use the sum of the births in place of  the deaths since the births or deaths of every cycle are the same order, as shown in Fig.~\ref{bc_native} and ~\ref{bc_misfold}. However, if we analyze  more complex molecules, the analysis may depend on the TFV definition. Chignolin is a small molecule with only one $\beta$-sheet and no tertiary structure. If we need to investigate a more complex protein with tertiary structures, then the TFV definition may affect the analysis results. Another improvement could be achieved by the selection of the TFV analysis method. In this study, we applied NMF to reduce the structure into two dimensions; however, when analyzing more complex proteins, the reduced dimensions will increase, so requiring careful determination of the appropriate reduced space dimension. In this case, other analysis method may be more appropriate. Once the TFV is calculated, we can apply several data mining and time-series analysis methods, such as hierarchical clustering, PCA, Fourier analysis, RMA, and ICA. In time-series analysis, Markov models are also promising, as it is a powerful tool for time-series analysis of protein dynamics\cite{Fujisaki2018}. The difficulty in applying Markov modeling lies in the definition of the states. In our study, we identified the misfolded and folded states, however, it is difficult to identify the boundary between them. Application of a hidden Markov model(HMM)\cite{Baum1966} may solve this problem as this method classifies each state automatically. Moreover, we can also apply text-mining methods. In our approach, an edge is regarded as a "term" to describe the protein shape. In this analogy, the set of generators is a document that describes the protein shape, and the volume optimal cycles are the sentences in the document. Therefore, we will be able to use text-mining methods such as topic models, term network analysis, and deep learning.

The method we developed here is applicable not only to protein folding but also to other problems in physics, chemistry, and engineering. 
For example, we could capture protein binding with small molecules, which will contribute to new drug development. Another interesting application is active matter dynamics, such as schools of fish or flocks of birds. Although the active matter dynamics is keenly studied, quantitatively analyzing the shapes of clusters of active matter is difficult. Our method will provide insights regarding this problem.

\section*{Author Contribution}
T. I. performed research. I. O. and Y. H. contributed to the development of analytical tools. T. I, I. O and Y. H. wrote the paper. 

\section*{Acknowledgement}
The authors thank to Hiroya Nakao, Hiromichi Suetani, Takenobu Nakamura, Kenji Fukumizu for their helpful comments. We would like to thank Editage (www.editage.com) for English language editing. This work was financially supported by JST CREST grant number JPMJCR15D3, Japan.

\bibliography{reference}
\end{document}